\documentclass[sigconf]{acmart}

\AtBeginDocument{%
  }
\usepackage{lipsum}  

\usepackage{enumitem}
\usepackage{multirow}
\usepackage{kotex}
\usepackage{booktabs}
\usepackage{makecell}
\newcommand{\proposed}{\textsf{LLM-SRec}}

\copyrightyear{2025}
\acmYear{2025}
\setcopyright{cc}
\setcctype{by}
\acmConference[KDD '25]{Proceedings of the 31st ACM SIGKDD Conference on Knowledge Discovery and Data Mining V.2}{August 3--7, 2025}{Toronto, ON, Canada}
\acmBooktitle{Proceedings of the 31st ACM SIGKDD Conference on Knowledge Discovery and Data Mining V.2 (KDD '25), August 3--7, 2025, Toronto, ON, Canada}
\acmDOI{10.1145/3711896.3737035}
\acmISBN{979-8-4007-1454-2/2025/08}
\begin{document}


\title{Lost in Sequence: Do Large Language Models Understand Sequential Recommendation?}

\author{Sein Kim}
\authornote{Both authors contributed equally to this research.}
\email{rlatpdlsgns@kaist.ac.kr}
\affiliation{
\institution{KAIST}
\city{Daejeon}
\country{Republic of Korea}
}

\author{Hongseok Kang}
\authornotemark[1]
\email{ghdtjr0311@kaist.ac.kr}
\affiliation{
\institution{KAIST}
\city{Daejeon}
\country{Republic of Korea}
}

\author{Kibum Kim}
\email{kb.kim@kaist.ac.kr}
\affiliation{
\institution{KAIST}
\city{Daejeon}
\country{Republic of Korea}
}

\author{Jiwan Kim}
\email{kim.jiwan@kaist.ac.kr}
\affiliation{
\institution{KAIST}
\city{Daejeon}
\country{Republic of Korea}
}

\author{Donghyun Kim}
\email{amandus.kim@navercorp.com}
\affiliation{
\institution{NAVER Corperation}
\city{Seongnam}
\country{Republic of Korea}
}

\author{Minchul Yang}
\email{minchul.yang@navercorp.com}
\affiliation{
\institution{NAVER Corperation}
\city{Seongnam}
\country{Republic of Korea}
}

\author{Kwangjin Oh}
\email{kj.oh@navercorp.com}
\affiliation{
\institution{NAVER Corperation}
\city{Seongnam}
\country{Republic of Korea}
}

\author{Julian McAuley}
\email{jmcauley@ucsd.edu}
\affiliation{
\institution{University of California San Diego}
\city{California}
\country{USA}
}

\author{Chanyoung Park}
\authornote{Corresponding author.}
\email{cy.park@kaist.ac.kr}
\affiliation{
\institution{KAIST}
\city{Daejeon}
\country{Republic of Korea}
}
\renewcommand{\shortauthors}{Sein Kim et al.}

\begin{abstract}
Large Language Models (LLMs) have recently emerged as promising tools for recommendation thanks to their advanced textual understanding ability and context-awareness. Despite the current practice of training and evaluating LLM-based recommendation (LLM4Rec) models under a sequential recommendation scenario, we found that whether these models understand the sequential information inherent in users’ item interaction sequences has been largely overlooked. In this paper, we first demonstrate through a series of experiments that existing LLM4Rec models do not fully capture sequential information both during training and inference. Then, we propose a simple yet effective LLM-based sequential recommender, called~\proposed, a method that enhances the integration of sequential information into LLMs by distilling the user representations extracted from a pre-trained CF-SRec model into LLMs.
Our extensive experiments show that~\proposed~enhances LLMs' ability to understand users' item interaction sequences, ultimately leading to improved recommendation performance.
Furthermore, unlike existing LLM4Rec models that require fine-tuning of LLMs, \proposed~achieves state-of-the-art performance by training only a few lightweight MLPs, highlighting its practicality in real-world applications. Our code is available at \url{https://github.com/Sein-Kim/LLM-SRec}.

\end{abstract}





\maketitle

\vspace{-1ex}
\section{Introduction}

Early efforts in LLM-based recommendation (LLM4Rec), such as TALLRec \cite{bao2023tallrec}, highlighted a gap between the capabilities of LLMs in text generation and sequential recommendation tasks, and proposed to address the gap by fine-tuning LLMs for sequential recommendation tasks using LoRA \cite{hu2022lora}. Subsequent studies, including LLaRA \cite{10.1145/3626772.3657690}, CoLLM \cite{zhang2023collm}, and A-LLMRec \cite{10.1145/3637528.3671931}, criticized the exclusive reliance of TALLRec on textual modalities, which rather limited its recommendation performance in warm scenarios (i.e., recommendation scenarios with abundant user-item interactions) \cite{zhang2023collm, 10.1145/3637528.3671931}. 
These methods transform item interaction sequences into text and provide them as prompts to LLMs \cite{10.1145/3626772.3657690} or align LLMs with a pre-trained Collaborative filtering-based sequential recommender (CF-SRec), such as SASRec \cite{kang2018self}, to incorporate the collaborative knowledge into LLMs \cite{10.1145/3637528.3671931}.
Despite the current practice of training and evaluating LLM4Rec models under a sequential recommendation scenario, we found that whether these models understand the sequential information inherent in users' item interaction sequences has been largely overlooked.
Hence, in this paper, we begin by conducting a series of experiments that are designed to investigate the ability of existing LLM4Rec models in understanding users' item interaction sequences (Sec.~\ref{sec: sequence exp}). More precisely, we compare four different LLM4Rec models (i.e., TALLRec, LLaRA, CoLLM, and A-LLMRec) with a CF-SRec model (i.e., SASRec). Our experimental results reveal surprising findings as follows:
\begin{enumerate}[leftmargin=0.5cm]
    \item \textbf{Training and Inference with Shuffled Sequences:} 
    {Randomly shuffling the order of items within a user's item interaction sequence breaks the sequential dependencies among items \cite{woolridge2021sequence, 10.1145/3640457.3688195}. Hence, we hypothesize that the performance of models that understand the sequential information inherent in a user's item interaction sequence would deteriorate when the sequence is disrupted. To investigate this, we conduct experiments under two different settings.
    First, we compare the performance of models that have been trained on the original sequences (i.e., non-shuffled sequences) and those trained on randomly shuffled sequences when they are evaluated on the same test sequences in which sequential information is present (i.e., non-shuffled test sequences). 
    Surprisingly, the performance of LLM4Rec models, even after being trained on shuffled sequences, is similar to the case when they are trained on the original sequences\footnote{{To address a potential concern that the moderate performance drop in LLM4Rec may be due to the prevalence of textual information over sequential data, we would like to emphasize that both types of information are indeed essential, as demonstrated in  Sec.~\ref{sec: case study}.
    On the other hand, disrupting the sequential information in users' item interaction sequences via shuffling still allows us to assess how effectively the models, particularly LLM4Rec, capture the sequential information, highlighting the importance of sequential information alongside textual data.}}, while SASRec trained with shuffled sequences shows significant performance degradation when tested on the original sequences. 
    Second, we perform inferences using shuffled sequences on the models that have been trained using the original sequences. Similar to our observations in the first experiment, we observed that LLM4Rec models exhibit minimal performance decline even when the sequences are shuffled during inference, while SASRec showed significant performance degradation. 
    In summary, these observations indicate that LLM4Rec models do not fully capture sequential information both during training and inference.}

    \item \textbf{Representation Similarity:} 
    In LLM4Rec models as well as in SASRec, representations of users are generated based on their item interaction sequences. Hence, we hypothesize that user representations obtained from a model that successfully captures sequential information in users' interaction sequences would greatly change when the input sequences are disrupted. To investigate this, we compute the similarity between user representations obtained based on the original sequences and those obtained based on shuffled sequences during inference.
    Surprisingly, the similarity is much higher for LLM4Rec models compared with that for SASRec, meaning that shuffling users' item interaction sequences has minimal impact on user representations of LLM4Rec models.
    This indicates again that LLM4Rec models do not fully capture sequential information.
\end{enumerate}

Motivated by the above findings, we propose a simple yet effective LLM-based sequential recommender, called \proposed, a method that enhances the integration of sequential information into LLMs. 
The main idea is to distill the user representations extracted from a pre-trained CF-SRec model into LLMs, so as to endow LLMs with the sequence understanding capability of the CF-SRec model.
Notably, our method achieves cost-efficient integration of sequential information without requiring fine-tuning of either the pre-trained CF-SRec models or the LLMs, effectively addressing the limitations of the existing LLM4Rec framework.
Our main contributions are summarized as follows:

\begin{table*}[h]
\small
\caption[Caption for LOF]{An example  prompt for various LLM4Rec models (Next Item Title Generation approach).
}
\vspace{-2ex}
\resizebox{0.95\linewidth}{!}{
\begin{tabular}{c|l|l|l}
\toprule
 \multicolumn{1}{c|}{}  & \multicolumn{1}{c|}{(a) \textbf{TALLRec}} & \multicolumn{1}{c|}{(b) \textbf{LLaRA}} & \multicolumn{1}{c}{(c) \textbf{CoLLM/A-LLMRec}} \\ \midrule\midrule
\multirow{3}{*}{\textbf{Inputs}} & This user has made a series of purchases& This user has made a series of purchases in the & This is user representation from recommendation models: \\ 
  & in the following order: (History Item List: & following order: (History Item List: [No.\# Time:& \textcolor{red}{[User Representation]}, and this user has made a series of purchases in\\  
 \multirow{2}{*}{$(\mathcal{P}^{u})$} &[No.\#  Time: YYYY/MM/DD Title: \textcolor{magenta}{Item Title}]).  & YYYY/MM/DD Title: \textcolor{magenta}{Item Title}, \textcolor{blue}{Item Embedding}]). & the following order: (History Item List: [No.\# Time: YYYY/\\
  &Choose one "Title" to recommend for this user&Choose one "Title" to recommend for this user to & MM/DD Title: \textcolor{magenta}{Item Title}, \textcolor{blue}{Item Embedding}]). Choose one "Title" to\\
 & to buy next from the following item "Title" set:& buy next from the following item "Title" set:  & recommend for this user to buy next from the following\\
  &  [Candidate \textcolor{magenta}{Item Titles}]. & [Candidate \textcolor{magenta}{Item Titles}, \textcolor{blue}{Item Embeddings}].&  item "Title" set: [Candidate \textcolor{magenta}{Item Titles}, \textcolor{blue}{Item Embeddings}].\\ \midrule
  
\textbf{Outputs}  &\multirow{2}{*}{Item Title} &\multirow{2}{*}{Item Title} &\multirow{2}{*}{Item Title} \\
$(\text{Text}(i_{n_u+1}^{(u)}))$  & & & \\ \bottomrule

\end{tabular}}
\label{tab title generation prompt}
\vspace{-1.5ex}
\end{table*}

\begin{table*}[h]
\small
\caption{An example prompt for various LLM4Rec models (Next Item Retrieval approach).
}
\vspace{-2ex}
\resizebox{0.95\linewidth}{!}{
\begin{tabular}{c|l|l|l}
\toprule
 \multicolumn{1}{c|}{}  & \multicolumn{1}{c|}{(a) \textbf{TALLRec}} & \multicolumn{1}{c|}{(b) \textbf{LLaRA/\proposed~(Ours)}} & \multicolumn{1}{c}{(c) \textbf{CoLLM/A-LLMRec}} \\ \midrule\midrule
\multirow{3}{*}{\textbf{User}} & This user has made a series of purchases& This user has made a series of purchases in the & This is user representation from recommendation models: \\ 
  & in the following order: (History Item List: & following order: (History Item List: [No.\# Time:& \textcolor{red}{[User Representation]}, and this user has made a series of purchases in\\  
  \multirow{2}{*}{$(\mathcal{P}^u_{\mathcal{U}})$}&[No.\#  Time: YYYY/MM/DD Title: \textcolor{magenta}{Item Title}]).  & YYYY/MM/DD Title: \textcolor{magenta}{Item Title}, \textcolor{blue}{Item Embedding}]). & the following order: (History Item List: [No.\# Time: YYYY/\\
  &Based on this sequence of purchases, generate & Based on this sequence of purchases, generate & MM/DD Title: \textcolor{magenta}{Item Title}, \textcolor{blue}{Item Embedding}]). Based on this\\
 &user representation token: \textcolor{brown}{[UserOut]}. & user representation token: \textcolor{brown}{[UserOut]}. &  sequence of purchases and user representation, generate\\
  &  & &  user representation token: \textcolor{brown}{[UserOut]}.\\ \midrule

    \textbf{Item} & The item title is as follows: "Title": \textcolor{magenta}{Item Title}, then & \multicolumn{2}{l}{The item title and item embedding are as follows: "Title": \textcolor{magenta}{Item Title}, \textcolor{blue}{Item Embedding}, then generate item representation} \\
    $(\mathcal{P}^i_{\mathcal{I}})$&generate item representation token: \textcolor{violet}{[ItemOut]}. & \multicolumn{2}{l}{token: \textcolor{violet}{[ItemOut]}} \\ \bottomrule
\end{tabular}}
\label{tab next item retrieval prompt}
\vspace{-1.5ex}
\end{table*}

\begin{itemize}[leftmargin=0.5cm]
\item We show that existing LLM4Rec models, although specifically designed for sequential recommendation, fail to effectively leverage the sequential information inherent in users' item interaction sequences.
\item We propose a simple and cost-efficient method that enables LLMs to capture the sequential information inherent in users' item interaction sequences for more effective recommendations.
\item {Our extensive experiments show that \proposed~outperforms existing LLM4Rec models by effectively capturing sequential dependencies. Furthermore, the results validate the effectiveness of transferring pre-trained sequential information through distillation method, across various experimental settings.}
\end{itemize}

\section{Do Existing LLM4Rec Models Understand Sequences?}
\label{sec sec2}

\subsection{Preliminaries}
\label{sec: problem setup}
\subsubsection{Definition of Sequential Recommendation in CF-SRec.}
Let $\mathcal{U} = \{u_1, u_2, \ldots, u_{|\mathcal{U}|}\}$ represent the set of users, and $\mathcal{I} = \{i_1, i_2, \ldots,$ $i_{|\mathcal{I}|}\}$ represent the set of items.
For a user $u \in \mathcal{U}$, $\mathcal{S}_u = (i_1^{(u)}, \ldots, i_t^{(u)},$ $\ldots, i_{n_u}^{(u)})$ denotes the item interaction sequence, where $i_t^{(u)} \in \mathcal{I}$ is the item that $u$ interacted with at time step $t$, and $n_u$ is the length of user $u$'s item interaction sequence.
Given the interaction history $\mathcal{S}_u$ of user $u$, the goal of sequential recommendation is to predict the next item that user $u$ will interact with at time step $n_{u}+1$ as $p(i_{n_u+1}^{(u)}\mid \mathcal{S}_u)$.
\subsubsection{LLM for Sequential Recommendation}

Note that existing LLM4Rec models can be largely categorized into the following two approaches: \textit{Generative Approach (}\textit{i.e., Next Item Title Generation}) \cite{10.1145/3637528.3671931,10.1145/3626772.3657690, hou2024large} and \textit{Retrieval Approach} (\textit{i.e., Next Item Retrieval}) \cite{geng2022recommendation, li2023e4srec}. 
In the Next Item Title Generation approach, a user's item interaction sequence and a list of candidate items are provided as input prompts to LLMs after which the LLMs \textit{generate} one of the candidate item titles as a recommendation. Meanwhile, the Next Item Retrieval approach extracts user and candidate item representations from the LLMs and \textit{retrieves} one of the candidate items whose similarity with the user representation is the highest.
Note that although existing LLM4Rec models have typically been proposed based on only one of the two approaches, we apply both approaches to each LLM4Rec baseline to conduct more comprehensive analyses on whether existing LLM4Rec models understand the sequential information inherent in users' item interaction sequences.

\smallskip
\noindent\textbf{1) Generative Approach (Next Item Title Generation).}
LLM4Rec models designed for Next Item Title Generation perform recommendations using instruction-based prompts as shown in Table~\ref{tab title generation prompt}. For a user $u$, the candidate item set of user $u$ is represented as $\mathcal{C}_u = \left\{i^{(u)}_{n_u+1}\right\} \cup \mathcal{N}_u$, where $\mathcal{N}_u = \text{RandomSample}(\mathcal{I}\backslash (\mathcal{S}_u \cup \left\{i^{(u)}_{n_u+1}\right\}) , m)$ is a negative item set for user $u$, and $m = |\mathcal{N}_u|$ is the number of negative items.
{Based on the item interaction sequence of user $u$, i.e., $\mathcal{S}_u$, and the candidate item set, i.e., $\mathcal{C}_u$, we write the input prompt $\mathcal{P}^u$ following the format shown in Table~\ref{tab title generation prompt}. 
Note that we introduce two projection layers, i.e., $f_{\mathcal{I}}$ and $f_{\mathcal{U}}$, each of which is used to project item embeddings and user representations extracted from a pre-trained CF-SRec into LLMs, respectively.
Following the completed prompts shown in Table~\ref{tab title generation prompt}, LLMs are trained for the sequential recommendation task through the Next Item Title Generation approach.
Note that TALLRec, LLaRA, and CoLLM use LoRA \cite{hu2022lora} to finetune LLMs aiming at learning the sequential recommendation task, while A-LLMRec only trains $f_{\mathcal{I}}$ and $f_{\mathcal{U}}$ without finetuning the LLMs with LoRA.
Please refer to the Appendix~\ref{app: next item title generation}.} for more details on the projection layers as well as prompt construction.

\smallskip
\noindent\textbf{2) Retrieval Approach (Next Item Retrieval).} 
As shown in Table ~\ref{tab next item retrieval prompt}, we use $\mathcal{P}^{u}_{\mathcal{U}}$ and $\mathcal{P}^{i}_{\mathcal{I}}$ to denote prompts for users and items, respectively.    
Unlike the Next Item Title Generation approach where LLMs directly generate the title of the recommended item, the Next Item Retrieval approach generates item recommendations by computing the recommendation scores between 
user representations and item embeddings.
{More precisely, it introduces learnable tokens, i.e., [UserOut] and [ItemOut], to aggregate information from user interaction sequences and items, respectively. The last hidden states associated with the [UserOut] and [ItemOut] are used as user representations and item embeddings, denoted $\mathbf{h}^{u}_{\mathcal{U}} \in \mathbb{R}^{l_{llm}}$ and $\mathbf{h}^{i}_{\mathcal{I}} \in \mathbb{R}^{l_{llm}}$, respectively, where $d_{llm}$ denotes the token embedding dimension of LLM. 
Please refer to Appendix ~\ref{app: next item retrieval}. for more details on how the user representations and item embeddings are extracted as well as the prompt construction for compared models.
Then, we compute the recommendation score between user $u$ and item $i$ as $s(u,i) = f_{\mathit{item}}(\mathbf{h}^{i}_{\mathcal{I}}) \cdot f_\mathit{user}(\mathbf{h}^u_{\mathcal{U}})^T$, where $f_\mathit{item}$ and $f_\mathit{user}$ are 2-layer MLPs, i.e., {$f_\mathit{item}, f_\mathit{user}:\mathbb{R}^{d_\mathit{llm}} \rightarrow \mathbb{R}^{d'}$}. Finally, the Next Item Retrieval loss is defined as follows:
\begin{equation}
\small
    \mathcal{L}_\text{Retrieval} = -\underset{u\in\mathcal{U}}{\mathbb{E}}[\text{log}\frac{e^{s(u,i^{(u)}_{n_u+1})}}{\sum_{k\in\mathcal{C}_u} e^{s(u,k)}}]
    \label{Eq retrieval}
\end{equation}
All models are trained using the $\mathcal{L}_\text{Retrieval}$ loss. Specifically, the set of MLPs (i.e., $f_{\mathcal{I}}, f_{\mathcal{U}}, f_\mathit{item}, f_\mathit{user}$, and two token embeddings (i.e., $\text{[ItemOut]}, \text{[UserOut]}$) are trained, while the LLM is fine-tuned using the LoRA. In contrast, A-LLMRec does not fine-tune the LLM.




\smallskip
\noindent\textbf{Discussion regarding prompt design. }
It is important to highlight that the prompts in Table~\ref{tab title generation prompt} and Table~\ref{tab next item retrieval prompt} are designed to ensure that LLMs interpret the user interaction history as a sequential process. Specifically, we incorporate both the interaction number and the actual timestamp of each interaction. Additionally, when shuffling the sequence, we only rearrange the item titles and embeddings while keeping the position of interaction number and timestamp unchanged. We considered that this choice is the most effective, as it allows us to maintain the integrity of the chronological order while still testing the model’s ability to generalize across different item sequences.

\vspace{-1ex}
\subsection{Evaluation Protocol}
In our experiments on LLMs' sequence comprehension, we employed the leave-last-out evaluation method (i.e., next item recommendation task) \cite{kang2018self,sun2019bert4rec,10.1145/3159652.3159656}. For each user, we reserved the last item in their behavior sequence as the test data, used the second-to-last item as the validation set, and utilized the remaining items for training. 
The candidate item set (i.e., test set) for each user in the title generation task is generated by randomly selecting 19 non-interacted items along with 1 positive item following existing studies \cite{zhang2023collm, 10.1145/3637528.3671931}. Similarly, for the next item retrieval task, following common strategies \cite{sun2019bert4rec,kang2018self}, we randomly select 99 non-interacted items along with 1 positive item as the candidate item set (i.e., test set) for each user. 

\vspace{-1ex}
\subsection{Preliminary Analysis}
\label{sec: sequence exp}
In this section, we conduct experiments to investigate the ability of LLM4Rec in understanding users’ item interaction sequences by comparing four different LLM4Rec models (i.e., TALLRec, LLaRA, CoLLM, and A-LLMRec)\footnotemark~ with a CF-SRec model (i.e., SASRec). Note that our experiments are designed based on the assumption that randomly shuffling the order of items within a user's item interaction sequence breaks the sequential dependencies among items \cite{10.1145/3640457.3688195, woolridge2021sequence}. More precisely, we conduct the following two experiments: 1) Training (Sec.~\ref{sec: shuffled training}) and Inference (Sec.~\ref{sec: shuffled inference}) with shuffled sequences, and 2) Representation Similarity (Sec.~\ref{sec: Representation Similarity}). In the following, we describe details regarding the experimental setup and experimental results.



\footnotetext{Note that TALLRec and CoLLM are designed for binary classification (YES/NO) for a target item, while LLaRA and A-LLMRec generate the title of item to be recommended (i.e., Next Item Title Generation approach). To adapt these baselines to the Next Item Retrieval approach, we modified their setup to retrieve the target item from a provided candidate item set by using the prompts in Table \ref{tab next item retrieval prompt} and training with Equation \ref{Eq retrieval}.}

\begin{table}[t]
\caption{Performance (NDCG@10) of various models when trained with original sequences and shuffled sequences (Next Item Retrieval approach). Change ratio indicates the performance change of `Shuffle' compared with `Original'.
}
\vspace{-1ex}
\resizebox{0.85\linewidth}{!}{
\begin{tabular}{c|c||c|c|c}
\toprule
         &  & Scientific & Electronics & CDs    \\ \midrule \midrule
\multirow{3}{*}{SASRec} &Original    & 0.2918  & 0.2267  & 0.3451 \\ 
&\multirow{1}{*}{Shuffle} & 0.2688 & 0.2104 &0.3312\\ \cmidrule{2-5}
&\multirow{1}{*}{Change ratio} &  (-7.88\%) & (-7.19\%) &  (-4.03\%) \\
\midrule \midrule

\multirow{3}{*}{TALLRec} & Original   & 0.2585  & 0.2249 & 0.3100 \\
&\multirow{1}{*}{Shuffle} & 0.2579 & 0.2223  & 0.3003 \\ \cmidrule{2-5}
&\multirow{1}{*}{Change ratio} &  (-0.23\%) & (-1.16\%) &  (-3.13\%) \\
\midrule \midrule

\multirow{3}{*}{LLaRA} & Original     & 0.2844     & 0.2048  & 0.2464 \\ 
&\multirow{1}{*}{Shuffle} & 0.2921 & 0.2079  & 0.2695  \\ \cmidrule{2-5}
&\multirow{1}{*}{Change ratio} & (+2.71\%) & (+1.51\%) & (+9.38\%) \\
\midrule \midrule

\multirow{3}{*}{CoLLM} &Original & 0.3111 & 0.2565   & 0.3152 \\ 
&\multirow{1}{*}{Shuffle} & 0.3181 & 0.2636  & 0.3143  \\ \cmidrule{2-5}
&\multirow{1}{*}{Change ratio} & (+2.25\%) & (+2.77\%) & (-0.29\%) \\
\midrule \midrule

\multirow{3}{*}{A-LLMRec} & Original  & 0.2875  & 0.2791  & 0.3119 \\
&\multirow{1}{*}{Shuffle} & 0.2973  & 0.2741  & 0.3078  \\ \cmidrule{2-5}
&\multirow{1}{*}{Change ratio} & (+3.41\%) & (-1.79\%) & (-1.31\%) \\
\midrule \midrule

\multirow{3}{*}{\proposed} & Original  & 0.3388  & 0.3044  & 0.3809 \\
&\multirow{1}{*}{Shuffle} & 0.3224  & 0.2838  & 0.3614  \\ \cmidrule{2-5}
&\multirow{1}{*}{Change ratio} & (-4.84\%) & (-6.77\%) & (-5.11\%) \\
\bottomrule

\end{tabular}}
\label{tab: shuffle train}
\vspace{-1ex}
\end{table}

\begin{table}[t]
\caption{Performance (HR@1) of various models when trained with original sequences and shuffled sequences (Next Item Title Generation approach).
}
\vspace{-1ex}
\resizebox{0.85\linewidth}{!}{
\begin{tabular}{c|c||c|c|c}
\toprule
         &  & Scientific & Electronics & CDs    \\ \midrule \midrule
\multirow{3}{*}{SASRec} &Original    & 0.3171  & 0.2390  & 0.3662 \\ 
&\multirow{1}{*}{Shuffle} & 0.2821 & 0.2158 &0.3386\\ \cmidrule{2-5}
&\multirow{1}{*}{Change ratio} &  (-11.04\%) & (-9.71\%) &  (-7.54\%) \\
\midrule \midrule

\multirow{3}{*}{TALLRec} & Original   & 0.2221  & 0.1787 & 0.2589 \\
&\multirow{1}{*}{Shuffle} & 0.2181 & 0.1815  & 0.2728 \\ \cmidrule{2-5}
&\multirow{1}{*}{Change ratio} &  (-1.81\%) & (+1.57\%) &  (+5.37\%) \\
\midrule \midrule

\multirow{3}{*}{LLaRA} & Original     & 0.3022     & 0.2616  & 0.3142 \\ 
&\multirow{1}{*}{Shuffle} & 0.2996 & 0.2650  & 0.3530  \\ \cmidrule{2-5}
&\multirow{1}{*}{Change ratio} & (-0.86\%) & (+1.30\%) & (+12.35\%) \\
\midrule \midrule

\multirow{3}{*}{CoLLM} &Original & 0.3010 & 0.2311   & 0.3447 \\ 
&\multirow{1}{*}{Shuffle} & 0.3165 & 0.2323  & 0.3763  \\ \cmidrule{2-5}
&\multirow{1}{*}{Change ratio} & (+5.15\%) & (+0.52\%) & (+9.17\%) \\
\midrule \midrule

\multirow{3}{*}{A-LLMRec} & Original  & 0.2804  & 0.2672  & 0.3319 \\
&\multirow{1}{*}{Shuffle} & 0.2796  & 0.2684  & 0.3528  \\ \cmidrule{2-5}
&\multirow{1}{*}{Change ratio} & (-0.29\%) & (+0.45\%) & (+6.30\%) \\
\bottomrule

\end{tabular}}
\label{tab: shuffle train - title generation}
\vspace{-1.5ex}
\end{table}

\subsubsection{\textbf{Shuffled Training}}
\label{sec: shuffled training}
We hypothesize that the performance of models trained on sequences in which meaningful sequential information is removed would deteriorate when evaluated on sequences in which sequential information is present (i.e., non-shuffled test sequences). To investigate this, we compare the performance of models that have been trained on the original sequence $\mathcal{S}_u$ (i.e., non-shuffled sequence) for each user $u$ and those trained on randomly shuffled sequences \cite{woolridge2021sequence}, when they are evaluated on the same non-shuffled test sequences. 
Note that users' item interaction sequences are shuffled only once before training begins, rather than at every epoch, to eliminate unintended augmentation effect \cite{takahagi-shinnou-2023-data}.

Table \ref{tab: shuffle train} and  Table \ref{tab: shuffle train - title generation} show the performance of various models when adapted to the Next Item Retrieval approach and the Next Item Title Generation approach, respectively.
We have the following observations:
1) CF-SRec (i.e., SASRec), suffers from substantial performance degradation when the training sequences are shuffled as expected, whereas LLM4Rec models generally exhibit minimal changes in performance. This indicates that LLMs struggle to leverage sequential information, as eliminating original sequential dependencies through shuffling does not significantly impact their performance.
2) Some LLM4Rec models even show improved results despite being trained with shuffled sequences. We conjecture that in some cases random shuffling of interaction sequences during training can introduce short-term co-occurrence patterns that may coincidentally lead to improved performance. Combined with the fact that LLMs struggle to capture long-term dependencies \cite{liu2024blockwise}, we argue that LLM4Rec models struggle to capture the sequential dependencies within the interaction sequence. 

\vspace{-1ex}
\subsubsection{\textbf{Shuffled Inference}}
\label{sec: shuffled inference}
We hypothesize that the performance of models that understand the sequential information inherent in a user’s item interaction sequence would deteriorate when the sequence is disrupted. To investigate this, we perform inference using shuffled test sequences with the models that have been trained using the original
sequences $\mathcal{S}_u$ (i.e., non-shuffled sequence).
That is, unlike in Sec.~\ref{sec: shuffled training}, we shuffle the test sequences during inference rather than training sequences.
It is important to note that the assumption of this experiment is that models trained with the original sequences indeed capture the sequential information.

\begin{figure}[t]
    \centering    \includegraphics[width=0.99\linewidth]{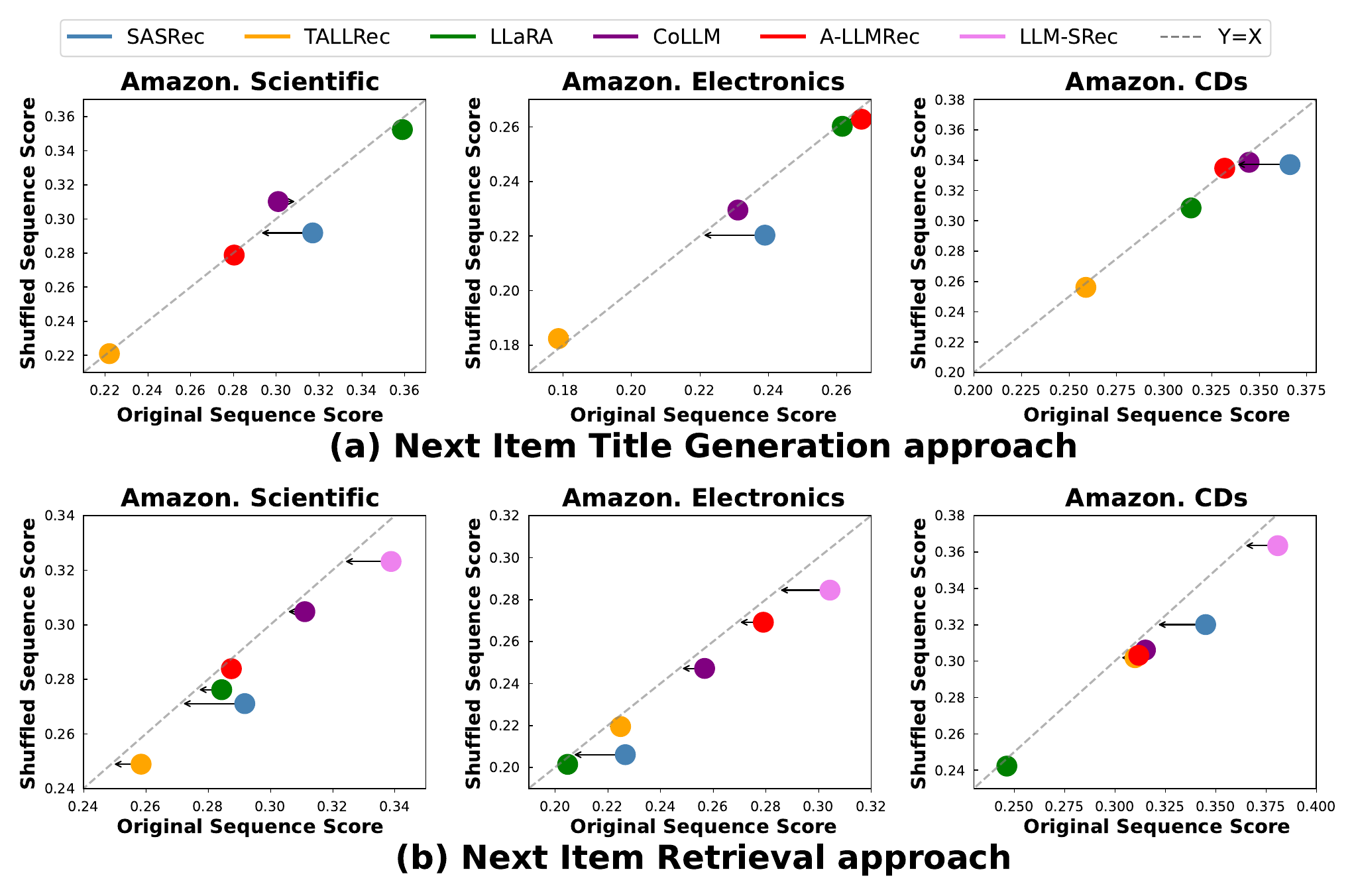}
    \vspace{-1ex}
    \caption
    {Performance of various models when tested with original sequences and shuffled sequences. (a) Next Item Title Generation approach (HR@1). (b) Next Item Retrieval approach (NDCG@10). "$\leftarrow$" indicates performance drop.}
    \label{fig: original_shuffle_perform_drop_all.pdf}
    \vspace{-2ex}
\end{figure}




{Figure \ref{fig: original_shuffle_perform_drop_all.pdf} (a) and (b) show the performance of various models when adapted to the Next Item Title Generation approach and the Next Item Retrieval approach, respectively.}
We have the following observations: 1) When the test sequences are shuffled, CF-SRec (i.e., SASRec) encounters a significant performance degradation as expected, whereas LLM4Rec remains relatively consistent (i.e., all circles except for SASRec are positioned near the $y=x$ in {Figure \ref{fig: original_shuffle_perform_drop_all.pdf} (a) and (b)).}
This implies that existing LLM4Rec models failed to capture the sequential information contained in the item interaction sequences.
2) Even though CoLLM and A-LLMRec leverage user representations derived from CF-SRec, which is solely trained based on the item interaction sequence, they fail to effectively capture the sequential information. This indicates the need for carefully distilling the user representations extracted from a pre-trained CF-SRec into LLMs.




\begin{table}[t]
\caption{Cosine similarity between user representations obtained based on original sequences and
those obtained based on shuffled sequences during inference (The models are trained on the original sequences).}
\vspace{-1ex}
\resizebox{0.8\linewidth}{!}{
\begin{tabular}{c||c|c|c|c}
\toprule
         & Movies & Scientific & Electronics & CDs    \\ \midrule\midrule
SASRec   & 0.6535 & 0.7375     & 0.7083      & 0.7454 \\ \midrule
TALLRec  & 0.9731 & 0.9326     & 0.9678      & 0.9570 \\ \midrule
LLaRA    & 0.9639 & 0.9424     & 0.9800      & 0.9624 \\ \midrule
CoLLM    & 0.9067 & 0.9263     & 0.8921      & 0.9526 \\ \midrule
A-LLMRec & 0.8872 & 0.8911     & 0.8623      & 0.8706 \\ \midrule
\proposed & 0.6128 & 0.7852     & 0.7393      & 0.8589 \\ \bottomrule
\end{tabular}}
\label{tab: user similarity}
\vspace{-2.25ex}
\end{table}

\subsubsection{\textbf{Representation Similarity}}
\label{sec: Representation Similarity}
In LLM4Rec models as well as in SASRec, representations of users are generated based on their item interaction sequences. Hence, we hypothesize that the user representations obtained from models that capture sequential information in a user’s item interaction sequence would change when the input sequence is disrupted. To investigate this, we compute the cosine similarity between user representations obtained based on the original sequences and those obtained based on shuffled sequences during inference. 

As shown in Table \ref{tab: user similarity}, we observe that the similarity is much higher for LLM4Rec models compared with that of CF-SRec, i.e., SASRec, indicating that LLM4Rec models are less effective than CF-SRec in capturing and reflecting changes in users' item interaction sequences.
It is worth noting that among LLM4Rec models, CoLLM and A-LLMRec exhibit relatively low similarity. This is attributed to the fact that they utilize user representations extracted from a pre-trained CF-SRec in their prompts, unlike TALLRec which only uses text, or LLaRA which uses text and item embeddings. 
This implies that incorporating user representations enhances the ability to effectively model sequential information. However, CoLLM and A-LLMRec still exhibit higher similarity values compared to SASRec, and based on the results of previous experiments (i.e., Sec. \ref{sec: shuffled training} and Sec. \ref{sec: shuffled inference}, it is evident that they have yet to fully comprehend the sequential patterns.


\vspace{-1ex}
\section{METHODOLOGY: \proposed}
\label{sec: method}
In this section, we propose ~\proposed, a novel and simplistic LLM4Rec framework designed to enable LLMs to effectively utilize sequential information inherent in users' item interaction sequences. 
It is important to note that among the two prominent approaches for LLM4Rec, we employ the Next Item Retrieval approach (i.e.,~\proposed~is trained with Equation \ref{Eq retrieval}) due to well-known drawbacks of the Next Item Title Generation approach, 
i.e. restrictions on the number of candidate items \cite{10.1145/3637528.3671931} and the existence of position bias with candidate items \cite{hou2024large}.
In the following, we explain two additional losses aiming at: (1) distilling the sequential information extracted from a pre-trained CF-SRec to LLMs (Sec. \ref{sec: distill}), and (2) preventing the over-smoothing problem during distillation (Sec. \ref{sec: uniform}).
Figure \ref{fig: framework} shows the overall framework of ~\proposed.


\begin{figure}[t]
    \centering
    \includegraphics[width=0.9\linewidth]{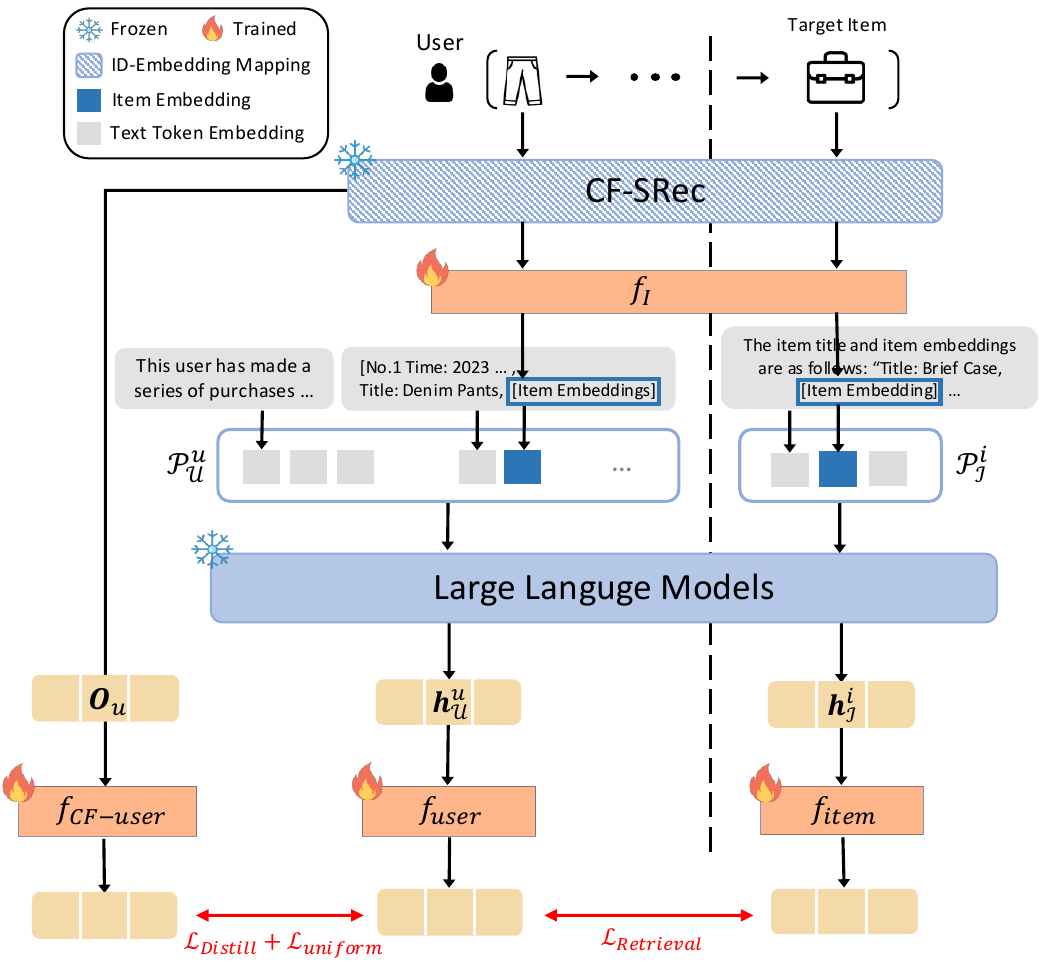}
    \vspace{-1.5ex}
\caption{Overall model architecture of ~\proposed.}
    \label{fig: framework}
    \vspace{-2ex}
\end{figure}

\vspace{-1ex}
\subsection{Distilling Sequential Information}
\label{sec: distill}
User representations from CF-SRec, derived solely from users' item interaction sequences, encapsulate rich sequential information crucial for sequential recommendation tasks. Despite the efforts of CoLLM and A-LLMRec trying to understand the sequential information by incorporating the user representations directly into LLM prompts, we observe that they still fail to do so (as shown in Sec.~\ref{sec sec2}). Therefore, in this paper, we propose a simple alternative approach to effectively incorporating the sequential information extracted from CF-SRec into LLMs. 
The main idea is to distill the sequential knowledge from pre-trained and frozen CF-SRec into LLMs.
More precisely, we simply match the user representation generated by a pre-trained CF-SRec, i.e., {$\mathbf{O}_u = \text{CF-SRec}(\mathcal{S}_u) \in \mathbb{R}^d$}, and that generated by LLMs, i.e., $\mathbf{h}^u$, as follows:
\begin{equation}
\small
    \mathcal{L}_\text{Distill} = \underset{u \in \mathcal{U}}{\mathbb{E}}[\textsf{MSE}(f_\mathit{CF-user}(\mathbf{O}_u), f_\mathit{user}(\mathbf{h}^u_{\mathcal{U}}))]
    \label{Eq distill}
\end{equation}
where $\textsf{MSE}$ is the mean squared error loss and both $f_\mathit{CF-user}$ and $f_\mathit{user}$ are 2-layer MLPs {, i.e., $f_\mathit{CF-user}: \mathbb{R}^d \rightarrow \mathbb{R}^{d'}$ and $f_\mathit{user}: \mathbb{R}^{d_{llm}} \rightarrow \mathbb{R}^{d'}$,} that are trainable.
This simple distillation framework enables LLMs to generate user representations that effectively capture and reflect the sequential information inherent in users' item interaction sequences.
The prompt used in ~\proposed~ is provided in Table \ref{tab next item retrieval prompt}, where user representations are derived from LLMs based on interacted item titles and collaborative information. {In Appendix \ref{app: prompt study}}, we empirically compare our prompt with another prompt that explicitly incorporates user representations, i.e., CoLLM and A-LLMRec, demonstrating the effectiveness of~\proposed~despite the absence of user representation in the prompt.
Additionally, Appendix \ref{app: contra distillation}, we conducted experiments using a contrastive learning approach as an objective for distillation instead of the MSE loss.

\vspace{-1ex}
\subsection{Preventing Over-smoothing}
\label{sec: uniform}
{Simply applying an MSE loss for distillation as in Equation \ref{Eq distill} can lead to the over-smoothing problem, i.e., two representations are highly similar, hindering LLMs from effectively learning sequential information. In extreme cases, $f_{\text{user}}$ and $f_{\mathit{CF-user}}$ could be trained to produce identical outputs \cite{10.1145/3637528.3671931, 10.1145/3627673.3679535}.} To mitigate this over-smoothing problem, we introduce a uniformity loss \cite{wang2020understanding, 10.1145/3627673.3679535, 10.1145/3534678.3539253} as follows:
\begin{equation}
\small
\begin{split}
    \mathcal{L}_\text{Uniform} &= 
    \underset{u \in \mathcal{U}}{\mathbb{E}}[\underset{u' \in \mathcal{U}}{\mathbb{E}} [e^{-2\|f_\mathit{CF-user}(\mathbf{O}_u) - f_\mathit{CF-user}(\mathbf{O}_{u'})\|^2_2}]]\\
    &+ \underset{u \in \mathcal{U}}{\mathbb{E}} [\underset{u' \in \mathcal{U}}{\mathbb{E}} [e^{-2\|f_\mathit{user}(\mathbf{h}^u_{\mathcal{U}}) - f_\mathit{user}(\mathbf{h}^{u'}_{\mathcal{U}})\|^2_2}]]
    \end{split}
    \label{Eq uniform}
\end{equation}
 The uniformity loss $\mathcal{L}_\text{Uniform}$ ensures that user representations of different users are uniformly distributed across the normalized feature space on the hypersphere, preserving both separation and informativeness.

\smallskip
\noindent\textbf{Final objective. }The final objective of~\proposed~is computed as the sum of the Next Item Retrieval loss (Equation \ref{Eq retrieval}), the distillation loss (Equation \ref{Eq distill}), and the uniformity loss (Equation \ref{Eq uniform}) as follows:
\begin{equation}
\label{eq final}
\small
    \mathcal{L} = \mathcal{L}_\text{Retrieval} + \mathcal{L}_\text{Distill} + \mathcal{L}_\text{Uniform}
\end{equation}
It is important to note that ~\proposed~ does not require any additional training for either the pre-trained CF-SRec or LLMs during its training process. Instead, ~\proposed~ only optimizes a small set of MLP layers (i.e., $f_{\mathcal{I}}, f_\mathit{user}$, $f_\mathit{CF-user}$, and $f_\mathit{item}$) and two LLM tokens (i.e., [UserOut] and [ItemOut]). Therefore, ~\proposed~ achieves faster training and inference time compared to existing LLM4Rec baselines, including TALLRec, LLaRA, CoLLM, and A-LLMRec, results are described in Sec. \ref{exp: time efficiency}. Furthermore, for training efficiency, we only consider the last item in $\mathcal{S}_u$ for each user $u$ to minimize Equation~\ref{eq final} \cite{10.1145/3637528.3671931}. That is, for each user $u$, we predict the last item $i_{n_u}^{(u)}$ given the sequence $(i_1^{(u)}, \ldots, i_t^{(u)}, \ldots, i_{n_u-1}^{(u)})$. 
In Appendix~\ref{app: autoregressive}, we also show the results when considering all items, i.e., auto-regressive learning.

\smallskip
\noindent\textbf{Discussion on the Efficiency of~\proposed. }
Last but not least, unlike existing LLM4Rec models such as TALLRec, LLaRA, and CoLLM, all of which require fine-tuning of LLMs, ~\proposed~eliminates the need for additional training or fine-tuning on either the pre-trained CF-SRec or the LLMs. Despite its simplicity, ~\proposed~ is highly effective in equipping LLMs with the ability to comprehend sequential information, making it lightweight but effective for sequential recommendation tasks.

\section{Experiments}
\noindent\textbf{Datasets.} 
We conduct experiments on four Amazon datasets \cite{hou2024bridging}, i.e., Movies, Scientific, Electronics, and CDs. Following prior studies \cite{kang2018self, sun2019bert4rec}, we use five-core datasets consisting of users and items with a minimum of five interactions each. The statistics for each dataset after preprocessing are summarized in Table ~\ref{tab dataset} in Appendix ~\ref{app: dataset}.

\smallskip
\noindent\textbf{Baselines.} We compare three groups of models as our baselines: models that use only interaction sequences (CF-SRec: GRU4Rec \cite{hidasi2015session}, BERT4Rec \cite{sun2019bert4rec}, NextItNet \cite{yuan2019simple}, SASRec \cite{kang2018self}), Language Model based models (LM-based: CTRL \cite{li2023ctrl}, RECFORMER \cite{li2023text}), and Large Language Model based models (LLM4Rec: TALLRec \cite{bao2023tallrec}, LLaRA \cite{10.1145/3626772.3657690}, CoLLM \cite{zhang2023collm}, A-LLMRec \cite{10.1145/3637528.3671931}). For fair comparisons, we implemented all LLM4Rec baselines with Next Item Retrieval approach. Details regarding the baselines are provided in Appendix~\ref{app: baseline}.

\smallskip
\noindent\textbf{Evaluation Protocol.}
We employ the leave-last-out strategy \cite{kang2018self} for evaluation, where the most recent item in the user interaction sequence is used as the test item, the second most recent item as the validation item, and the remaining sequence for training. To evaluate the performance of sequential recommendation, we add 99 randomly selected non-interacted items to the test set, ensuring that each user's test set consists of one positive item and 99 negative items. Evaluation is conducted using two widely adopted metrics: Normalized Discounted Cumulative Gain (NDCG@N) and Hit Ratio (HR@N), with N set to 10 and 20.

\smallskip
\noindent\textbf{Implementation Details.}
For fair comparisons, we adopt pre-trained LLaMA 3.2 (3B-Instruct) as the backbone LLM for all LLM4Rec baselines (i.e., TALLRec, CoLLM, LLaRA, and A-LLMRec) including~\proposed. 
Similarly, SASRec serves as the pre-trained CF-SRec for CoLLM, LLaRA, A-LLMRec, and ~\proposed. Please refer to the Appendix \ref{app: implementation details} for more details regarding the hyper-parameters and train settings.


\begin{table*}[t]
\caption{Overall model performance. The best performance is denoted in bold.}
\vspace{-2ex}
\resizebox{0.875\linewidth}{!}{
\begin{tabular}{c|c||cccc||cc||ccccc}
\toprule
\multirow{2}{*}{Dataset} & \multirow{2}{*}{Metric} & \multicolumn{4}{c||}{CF-SRec}                                                           & \multicolumn{2}{c||}{LM-based}                                   & \multicolumn{5}{c}{LLM4Rec}                                                                                                      \\ \cmidrule{3-13} 
&                   & \multicolumn{1}{c|}{GRU4Rec} & \multicolumn{1}{c|}{BERT4Rec} & \multicolumn{1}{c|}{NextItNet} & SASRec & \multicolumn{1}{c|}{CTRL} & RECFORMER & \multicolumn{1}{c|}{TALLRec} & \multicolumn{1}{c|}{LLaRA} & \multicolumn{1}{c|}{CoLLM} & \multicolumn{1}{c|}{A-LLMRec} & \proposed \\ \midrule\midrule
\multirow{4}{*}{Movies} & NDCG@10           & \multicolumn{1}{c|}{0.3152}        & \multicolumn{1}{c|}{0.2959}         & \multicolumn{1}{c|}{0.2538}          & 0.3459 & \multicolumn{1}{c|}{0.2785} &   0.2068  & \multicolumn{1}{c|}{0.1668}        & \multicolumn{1}{c|}{0.1522}      & \multicolumn{1}{c|}{0.3223}      & \multicolumn{1}{c|}{0.3263}         &  \textbf{0.3560} \\ \cline{2-13} 
  & NDCG@20           & \multicolumn{1}{c|}{0.3494}        & \multicolumn{1}{c|}{0.3303}         & \multicolumn{1}{c|}{0.2879}          &  0.3745 & \multicolumn{1}{c|}{0.3099} &    0.2337  & \multicolumn{1}{c|}{0.2126}        & \multicolumn{1}{c|}{0.1944}      & \multicolumn{1}{c|}{0.3577}      & \multicolumn{1}{c|}{0.3629}         & \textbf{0.3924}  \\ \cline{2-13} 
  & HR@10             & \multicolumn{1}{c|}{0.4883}        & \multicolumn{1}{c|}{0.4785}         & \multicolumn{1}{c|}{0.4221}          & 0.5180  & \multicolumn{1}{c|}{0.4264} &    0.3569  & \multicolumn{1}{c|}{0.3234}        & \multicolumn{1}{c|}{0.2914}      & \multicolumn{1}{c|}{0.5089}      & \multicolumn{1}{c|}{0.5127}         & \textbf{0.5569} \\ \cline{2-13} 
  & HR@20             & \multicolumn{1}{c|}{0.6245}        & \multicolumn{1}{c|}{0.6213}         & \multicolumn{1}{c|}{0.5522}          &  0.6310 & \multicolumn{1}{c|}{0.5429} &    0.5264  & \multicolumn{1}{c|}{0.5060}        & \multicolumn{1}{c|}{0.4599}      & \multicolumn{1}{c|}{0.6491}      & \multicolumn{1}{c|}{0.6577}         & \textbf{0.7010} \\ \midrule\midrule
\multirow{4}{*}{Scientific} & NDCG@10           & \multicolumn{1}{c|}{0.2642}        & \multicolumn{1}{c|}{0.2576}         & \multicolumn{1}{c|}{0.2263}& 0.2918 & \multicolumn{1}{c|}{0.2152} &   0.2907    & \multicolumn{1}{c|}{0.2585}        & \multicolumn{1}{c|}{0.2844}      & \multicolumn{1}{c|}{0.3111}      & \multicolumn{1}{c|}{0.2875}         &   \textbf{0.3388}  \\ \cline{2-13} 
  & NDCG@20           & \multicolumn{1}{c|}{0.2974}        & \multicolumn{1}{c|}{0.2913}         & \multicolumn{1}{c|}{0.2657}          &  0.3245  & \multicolumn{1}{c|}{0.2520} &  0.3113  & \multicolumn{1}{c|}{0.3048}        & \multicolumn{1}{c|}{0.3265}      & \multicolumn{1}{c|}{0.3489}      & \multicolumn{1}{c|}{0.3246}         & \textbf{0.3758} \\ \cline{2-13} 
  & HR@10             & \multicolumn{1}{c|}{0.4313}        & \multicolumn{1}{c|}{0.4437}         & \multicolumn{1}{c|}{0.3908}          & 0.4691 & \multicolumn{1}{c|}{0.3520} &    0.4506  & \multicolumn{1}{c|}{0.4574}        & \multicolumn{1}{c|}{0.4993}      & \multicolumn{1}{c|}{0.5182}      & \multicolumn{1}{c|}{0.4957}         & \textbf{0.5532}  \\ \cline{2-13} 
  & HR@20             & \multicolumn{1}{c|}{0.5524}        & \multicolumn{1}{c|}{0.5822}         & \multicolumn{1}{c|}{0.5356}          & 0.5987   & \multicolumn{1}{c|}{0.4882} &    0.5710  & \multicolumn{1}{c|}{0.6276}        & \multicolumn{1}{c|}{0.6658}      & \multicolumn{1}{c|}{0.6676}      & \multicolumn{1}{c|}{0.6427}         & \textbf{0.6992} \\ \midrule\midrule
\multirow{4}{*}{Electronics} & NDCG@10           & \multicolumn{1}{c|}{0.2364}        & \multicolumn{1}{c|}{0.1867}         & \multicolumn{1}{c|}{0.1712}          &   0.2267 & \multicolumn{1}{c|}{0.1680} &  0.2032   & \multicolumn{1}{c|}{0.2249}        & \multicolumn{1}{c|}{0.2048}      & \multicolumn{1}{c|}{0.2565}      & \multicolumn{1}{c|}{0.2791}         & \textbf{0.3044} \\ \cline{2-13} 
  & NDCG@20           & \multicolumn{1}{c|}{0.2743}        & \multicolumn{1}{c|}{0.2172}         & \multicolumn{1}{c|}{0.2069}          &  0.2606 & \multicolumn{1}{c|}{0.2003} &   0.2441  & \multicolumn{1}{c|}{0.2670}        & \multicolumn{1}{c|}{0.2454}      & \multicolumn{1}{c|}{0.2948}      & \multicolumn{1}{c|}{0.3173}         & \textbf{0.3424} \\ \cline{2-13} 
  & HR@10             & \multicolumn{1}{c|}{0.3843}        & \multicolumn{1}{c|}{0.3325}         & \multicolumn{1}{c|}{0.3017}          &  0.3749  & \multicolumn{1}{c|}{0.2861} &    0.3586 & \multicolumn{1}{c|}{0.3802}        & \multicolumn{1}{c|}{0.3441}      & \multicolumn{1}{c|}{0.4236}      & \multicolumn{1}{c|}{0.4622}         & \textbf{0.4885} \\ \cline{2-13} 
  & HR@20             & \multicolumn{1}{c|}{0.5196}        & \multicolumn{1}{c|}{0.4740}         & \multicolumn{1}{c|}{0.4324}          & 0.5096  & \multicolumn{1}{c|}{0.4152} &    0.5213  & \multicolumn{1}{c|}{0.5476}        & \multicolumn{1}{c|}{0.5032}      & \multicolumn{1}{c|}{0.5741}      & \multicolumn{1}{c|}{0.6137}         & \textbf{0.6385} \\ \midrule\midrule
\multirow{4}{*}{CDs} & NDCG@10           & \multicolumn{1}{c|}{0.2155}        & \multicolumn{1}{c|}{0.3019}         & \multicolumn{1}{c|}{0.2207}          &  0.3451  & \multicolumn{1}{c|}{0.2968} &   0.3238   & \multicolumn{1}{c|}{0.3100}        & \multicolumn{1}{c|}{0.2464}      & \multicolumn{1}{c|}{0.3152}      & \multicolumn{1}{c|}{0.3119}         &  \textbf{0.3809}\\ \cline{2-13} 
  & NDCG@20           & \multicolumn{1}{c|}{0.2530}        & \multicolumn{1}{c|}{0.3386}         & \multicolumn{1}{c|}{0.2562}          & 0.3795  & \multicolumn{1}{c|}{0.3316} &    0.3642   & \multicolumn{1}{c|}{0.3493}        & \multicolumn{1}{c|}{0.2951}      & \multicolumn{1}{c|}{0.3557}      & \multicolumn{1}{c|}{0.3526}         & \textbf{0.4158}  \\ \cline{2-13} 
  & HR@10             & \multicolumn{1}{c|}{0.3712}        & \multicolumn{1}{c|}{0.5018}         & \multicolumn{1}{c|}{0.3842}          & 0.5278  & \multicolumn{1}{c|}{0.4574} &     0.5140  & \multicolumn{1}{c|}{0.5052}        & \multicolumn{1}{c|}{0.4665}      & \multicolumn{1}{c|}{0.5290}      & \multicolumn{1}{c|}{0.5300}         & \textbf{0.6085}  \\ \cline{2-13} 
  & HR@20             & \multicolumn{1}{c|}{0.5092}        & \multicolumn{1}{c|}{0.6605}         & \multicolumn{1}{c|}{0.5422}          & 0.6635  & \multicolumn{1}{c|}{0.5957} & 0.6739 & \multicolumn{1}{c|}{0.6633}        & \multicolumn{1}{c|}{0.6590}      & \multicolumn{1}{c|}{0.6895}      & \multicolumn{1}{c|}{0.6914}         & \textbf{0.7461} \\ \bottomrule
\end{tabular}}
\label{tab: overall performance}
\vspace{-1.5ex}
\end{table*}

\vspace{-1ex}
\subsection{Recommendation Performance Comparison}
\subsubsection{\textbf{Overall performance}}
\label{exp: overall}
Table \ref{tab: overall performance} presents the recommendation performance of ~\proposed~ and baselines on four datasets. From the results, we have the following observations:
1) \proposed~consistently outperforms existing LLM4Rec models. This result highlights the importance of distilling the sequential knowledge extracted from CF-SRec into LLMs.
2) \proposed~ outperforms CF-SRec based and LM-based models, suggesting that the reasoning ability and the pre-trained knowledge of LLMs significantly contribute to recommendation performance.
3) LLM4Rec models that utilize CF-SRec in their framework (i.e., LLaRA, CoLLM, and A-LLMRec) outperform TALLRec while being comparable to their CF-SRec backbone, i.e., SASRec. This indicates that while incorporating the CF knowledge extracted from a pre-trained CF-SRec is somewhat helpful, the lack of sequence understanding ability limits further improvements, even when using the LLMs. 
In summary, these findings emphasize the significance of seamless distillation of the sequential information extracted from a pre-trained CF-SRec into LLMs as in~\proposed.

\subsubsection{\textbf{Transition \& Non-Transition Sequences.}}
\label{sec transition_non-transition}

\begin{figure}[t]
    \centering
    \includegraphics[width=0.925\linewidth]{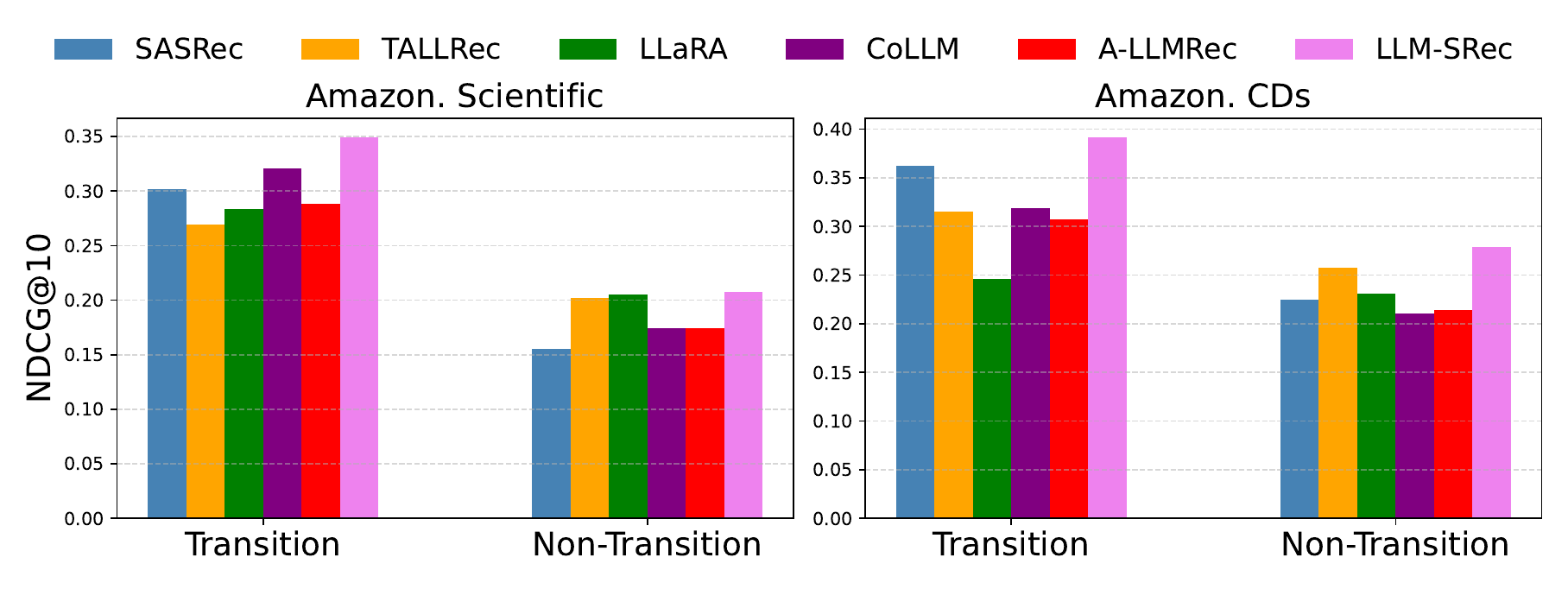}
    \vspace{-2ex}
    \caption
    {Performance with a varying degree of sequential information ("Transition Set" vs. "Non-Transition Set").}
    \label{fig: transition}
    \vspace{-3ex}
\end{figure}

To examine how the degree of sequential information contained in users' item interaction sequences influences the model performance, we categorized users based on the degree of sequential transitions in their interaction history\footnote{We count the number of unique transitions occurring between consecutive items, i.e., $i^{(u)}_{t} \rightarrow i^{(u)}_{t+1}$, within the sequence of user $u$, i.e., $\mathcal{S}^u$. Then, we sort users according to the transition score, i.e., $\textsf{t-score}^{u} = (\sum_{t=1}^{n_u-1}Count(i^{(u)}_{t} \rightarrow i^{(u)}_{t+1}))/(n^u-1)$. Users within the top-50\% are assigned to the "Transition Set", while the remaining users were assigned to the "Non-Transition Set." That is, users whose item interaction sequences exhibit sequential information are assigned to the "Transition Set".}.
We make the following observations from Figure \ref{fig: transition}. 
1) \proposed~outperforms all baselines, especially in the Transition Set, where sequential information is more abundant. This demonstrates that the distillation of sequence information enables the LLMs to comprehend and utilize sequential information inherent in users' interaction sequences.
2) 
The performance gap between LLM4Rec baselines and \proposed~is smaller in the Non-Transition Set compared with the Transition Set.
This indicates that existing LLM4Rec models lack the capability to effectively capture sequential dependencies among items and further emphasizes the importance of effective sequential modeling.

\begin{table}[h]
\caption{Performance on cross-domain scenarios (HR@10).}
\vspace{-1.5ex}
\resizebox{0.925\linewidth}{!}{
\begin{tabular}{c||c|c|c|c|c|c}
\toprule
& SASRec                                 & TALLRec                                & LLaRA                                  & CoLLM                                  & A-LLMRec                               & \proposed                               \\ \hline
Electronics $\rightarrow$ & \multirow{2}{*}{0.1002}                & \multirow{2}{*}{0.1214}                & \multirow{2}{*}{0.1225}                & \multirow{2}{*}{0.1232}                & \multirow{2}{*}{0.1262}                & \multirow{2}{*}{\textbf{0.1310}}                \\ 
Scientific                &                                        &                                        &                                        &                                        &                                        &                                        \\ \midrule\midrule
Electronics $\rightarrow$     & \multirow{2}{*}{0.0974}& \multirow{2}{*}{0.1132} & \multirow{2}{*}{0.1174} & \multirow{2}{*}{0.1152}& \multirow{2}{*}{0.1217}& \multirow{2}{*}{\textbf{0.1369}} \\ 
   CDs &             &                &                 &                &               &            \\ \bottomrule
\end{tabular}}
\label{tab: cross-domain}
\vspace{-1.5ex}
\end{table}

\begin{table*}[t]
\caption{Ablation studies on the components of~\proposed.}
\vspace{-2ex}
\resizebox{0.85\linewidth}{!}{
\begin{tabular}{c|c|c||cc||cc||cc||cc}
\toprule
\multirow{2}{*}{Row}&\multirow{2}{*}{Ablation} &\multirow{2}{*}{Train Set}  &\multicolumn{2}{c||}{Movies}& \multicolumn{2}{c||}{Scientific} & \multicolumn{2}{c||}{Electronics}& \multicolumn{2}{c}{CDs}\\ \cmidrule{4-11} 
& & & \multicolumn{1}{c|}{NDCG@10} & NDCG@20 & \multicolumn{1}{c|}{NDCG@10} &  NDCG@20 & \multicolumn{1}{c|}{NDCG@10} & NDCG@20 & \multicolumn{1}{c|}{NDCG@10} & NDCG@20   \\ \midrule\midrule
                  
\multirow{3}{*}{(a)}&\multirow{3}{*}{
\makecell{w.o. $\mathcal{L}_\text{Distill}$, \\$\mathcal{L}_\text{Uniform}$}} & Original  & \multicolumn{1}{c|}{0.3204}  & 0.3569   & \multicolumn{1}{c|}{0.3088}  & 0.3450 & \multicolumn{1}{c|}{0.2659}  & 0.3066 & \multicolumn{1}{c|}{0.2278}  & 0.2701  \\ 
& & \multirow{1}{*}{Shuffle} & \multicolumn{1}{c|}{0.3176}  & 0.3557 & \multicolumn{1}{c|}{0.3013}  & 0.3379 & \multicolumn{1}{c|}{0.2589}  & 0.2990   & \multicolumn{1}{c|}{0.2224} & 0.2649 \\ \cmidrule{3-11}
& &  \multirow{1}{*}{Change ratio} & \multicolumn{1}{c|}{(-0.87\%)}  & (-0.34\%)  & \multicolumn{1}{c|}{(-2.33\%)}  & (-2.06\%) & \multicolumn{1}{c|}{(-2.63\%)}  & (-2.48\%) & \multicolumn{1}{c|}{(-2.37\%)} & (-1.92\%)\\ 
\midrule\midrule

\multirow{3}{*}{(b)}&\multirow{3}{*}{w.o. $\mathcal{L}_\text{Uniform}$}  &  Original  & \multicolumn{1}{c|}{0.3339}  & 0.3700  & \multicolumn{1}{c|}{0.3283}  & 0.3653  & \multicolumn{1}{c|}{0.2895}  & 0.3285  & \multicolumn{1}{c|}{0.3622}  & 0.4013 \\
& & \multirow{1}{*}{Shuffle}         & \multicolumn{1}{c|}{0.3089}  & 0.3456 & \multicolumn{1}{c|}{0.3164}  & 0.3536 & \multicolumn{1}{c|}{0.2732}  & 0.3110 & \multicolumn{1}{c|}{0.3478}  & 0.3885 \\ \cmidrule{3-11}
& & \multirow{1}{*}{Change ratio}    & \multicolumn{1}{c|}{(-7.49\%)}  & (-6.59\%) & \multicolumn{1}{c|}{(-3.62\%)}  & (-3.20\%) & \multicolumn{1}{c|}{(-5.63\%)}  & (-5.33\%)  & \multicolumn{1}{c|}{(-3.98\%)} & (-3.19\%)\\
\midrule\midrule

\multirow{3}{*}{(c)}& \multirow{3}{*}{\proposed}  &  Original & \multicolumn{1}{c|}{\textbf{0.3560}}  & \textbf{0.3924} & \multicolumn{1}{c|}{\textbf{0.3388}}  & \textbf{0.3758} & \multicolumn{1}{c|}{\textbf{0.3044}}  & \textbf{0.3424} & \multicolumn{1}{c|}{\textbf{0.3809}}  & \textbf{0.4158} \\
& &  \multirow{1}{*}{Shuffle}   & \multicolumn{1}{c|}{0.3263}  & 0.3624& \multicolumn{1}{c|}{0.3224}  & 0.3591 & \multicolumn{1}{c|}{0.2838}  & 0.3210 & \multicolumn{1}{c|}{0.3614}        & 0.3981 \\ \cmidrule{3-11}
& &  \multirow{1}{*}{Change ratio}    & \multicolumn{1}{c|}{(-8.34\%)}  & (-7.65\%)& \multicolumn{1}{c|}{(-4.84\%)}  & (-4.44\%) & \multicolumn{1}{c|}{(-6.77\%)}  & (-6.25\%) & \multicolumn{1}{c|}{(-5.11\%)} & (-4.26\%)\\

\bottomrule
\end{tabular}}
\vspace{-1.5ex}
\label{tab: ablation}
\end{table*}

\begin{figure}[t]
    \centering
    \includegraphics[width=0.925\linewidth]{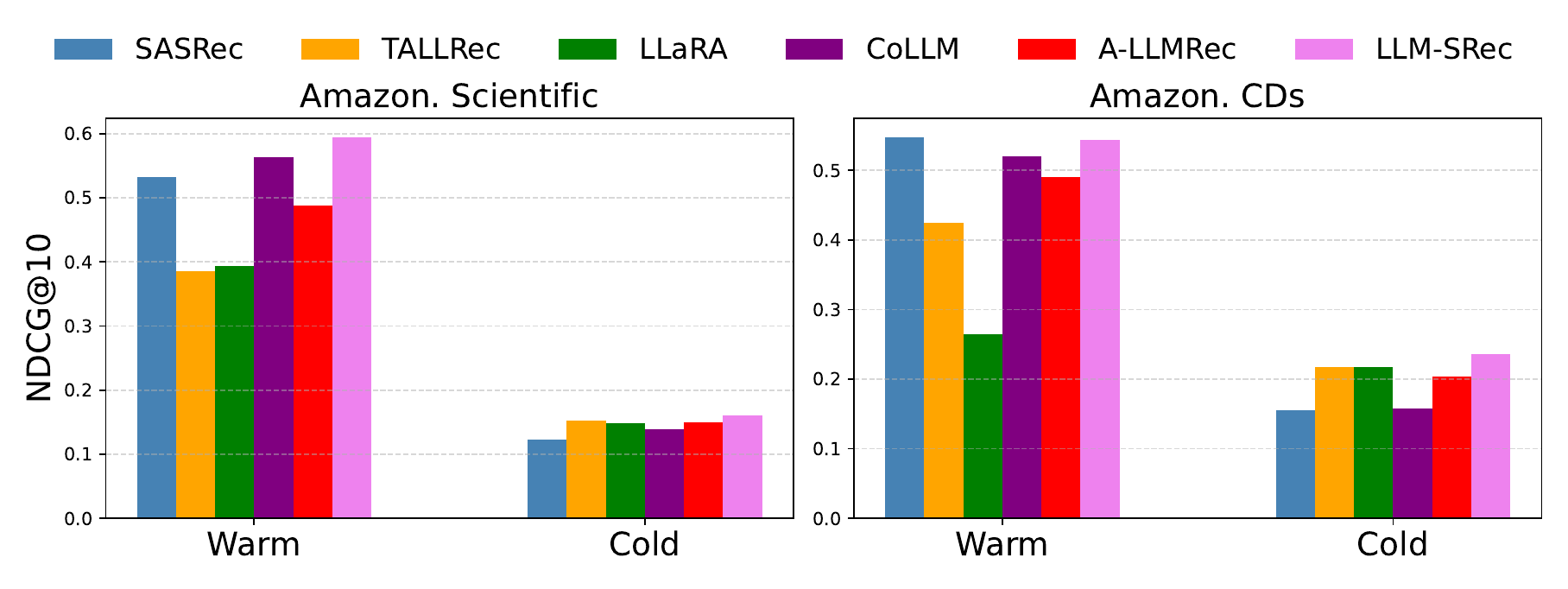}
    \vspace{-2ex}
    \caption
    {Performance on Warm/Cold item Scenarios.}
    \label{fig: warm cold}
    \vspace{-2.5ex}
\end{figure}

\vspace{-1ex}
\subsubsection{\textbf{Performance under Warm/Cold Scenarios.}}
In this section, we conduct experiments to examine how ~\proposed~performs in both warm and cold item settings.
Following the experimental setup of A-LLMRec \cite{10.1145/3637528.3671931}, items are labeled as ‘warm’ if they belong to the top 35\% in terms of the number of interactions with users, while those in the bottom 35\% are labeled as ‘cold’ items.
We have the following observations in Figure~\ref{fig: warm cold}: 1)~\proposed~ consistently achieves superior performance in both warm and cold scenarios, benefiting from its ability to capture the sequential information within item interaction sequences. 
Additionally, the performance in the cold setting shows that ~\proposed~ effectively leverages the generalizability of LLMs, utilizing pre-trained knowledge and textual understanding even though there is insufficient collaborative knowledge for cold items.
2) TALLRec, which relies solely on textual information, performs inferior in warm settings than LLM4Rec baselines (i.e., LLaRA, CoLLM, and A-LLMRec) that incorporate collaborative knowledge from CF-SRec. However, these models still underperform compared to ~\proposed, highlighting the necessity of modeling both collaborative and sequential information for effective LLM-based recommendation. 3) As expected, SASRec particularly struggles for cold items due to its exclusive reliance on the user-item interaction data. In contrast, LLM4Rec models, especially \proposed, leverage item textual information to mitigate the scarcity of interactions.

\vspace{-2.5ex}
\subsubsection{\textbf{Performance under Cross-domain Scenarios.}}
\label{sec: cross-domain}
{To further verify the generalizability of ~\proposed, we evaluate ~\proposed~ on the cross-domain scenarios, following the setting of A-LLMRec \cite{10.1145/3637528.3671931}, where the models are evaluated on datasets that have not been used for training. 
{Specifically, we pre-train the models on the Electronic dataset, as it contains the most users and items, and perform evaluations on the Scientific and CDs.}
As shown in Table~\ref{tab: cross-domain}, we observe that 
{1)~\proposed~ consistently outperforms all the baselines in the cross-domain scenarios. Leveraging the textual understanding of LLMs, ~\proposed~ extracts 
 textual information from unseen items which lack collaborative information. Additionally, by capturing sequential information from the source data, i.e., Electronics dataset, and aligning it with the textual understanding of LLMs, ~\proposed~generates high-quality user representations, enabling superior performance in cross-domain scenarios.}
2) While LLM4Rec baselines also address the issue of unseen items through LLM's textual understanding, they fail to generate user representations that capture sequential information, resulting in lower performance than ~\proposed. In contrast, CF-SRec struggles with unseen items due to the difficulty of generating collaborative knowledge of items, leading to inferior performance.}




\vspace{-1ex}
\subsection{Ablation Studies}
\label{exp: ablation}
In this section, we evaluate the contribution of each component~\proposed. To analyze not only the contribution of each component in terms of the final performance but also its impact on the sequence understanding ability, we conduct experiments under the setting described in Sec. \ref{sec: shuffled training}. In other words, we compare the performance of training with the original sequences versus training with shuffled sequences. 
Table \ref{tab: ablation} presents the following observations:
1) When both $\mathcal{L}_\text{Distill}$ and $\mathcal{L}_\text{Uniform}$ are present (i.e., vanilla \proposed~ in row (c)), the model consistently achieves the highest performance on the original sequence due to the benefits of sequence understanding ability. Furthermore, compared with its variants (i.e., row (c) vs (a,b)), the performance of vanilla \proposed~drops the most rapidly when shuffling is applied, ensuring that vanilla \proposed~indeed comprehends and effectively utilizes sequential information. 
2) In the absence of both $\mathcal{L}_\text{Distill}$ and $\mathcal{L}_\text{Uniform}$ (i.e., row (a)), where the sequential knowledge extracted from CF-SRec is not distilled to the LLMs, the model exhibits the lowest performance on the original sequence. Additionally, even when random shuffling is applied, the performance drop is minor. This result suggests that the model lacks sequence understanding capability without $\mathcal{L}_\text{Distill}$ and $\mathcal{L}_\text{Uniform}$. 
3) When only $\mathcal{L}_\text{Uniform}$ is removed, the model suffers from the over-smoothing problem as discussed in Sec. \ref{sec: uniform}, leading to lower performance compared to ~\proposed. However, since $\mathcal{L}_\text{Distill}$ is still present, we observe a significant performance drop when applying random shuffling. This once more indicates that the model is endowed with the sequence understanding ability with 
$\mathcal{L}_\text{Distill}$.

\begin{table}[t]
\caption{Train/Inference time comparison.}
\vspace{-1ex}
\resizebox{0.9\linewidth}{!}{
\begin{tabular}{c|c|c|c|c}
\toprule
 \multirow{2}{*}{}& \multicolumn{2}{c|}{Scientific} &\multicolumn{2}{c}{Electronics} \\ \cmidrule{2-5}
     & \multicolumn{1}{c|}{Train (min/epoch)}       & Inference (min)  & \multicolumn{1}{c|}{Train (min/epoch)} &  Inference (min) \\ \midrule
\midrule
TALLRec                  & \multicolumn{1}{c|}{194.43}         &  37.04  & \multicolumn{1}{c|}{236.73} &  29.04 \\ \midrule
LLaRA                    & \multicolumn{1}{c|}{202.20}         &  38.79 & \multicolumn{1}{c|}{241.17} & 30.62 \\ \midrule
CoLLM                    & \multicolumn{1}{c|}{214.12}         &      39.86 & \multicolumn{1}{c|}{251.51} & 32.58\\ \midrule
A-LLMRec                    & \multicolumn{1}{c|}{190.94}         &  35.01 & \multicolumn{1}{c|}{235.02} & 28.14 \\ \midrule\midrule
\proposed                & \multicolumn{1}{c|}{\textbf{185.91}}         &  \textbf{34.17}  & \multicolumn{1}{c|}{\textbf{218.21}} & \textbf{27.57}\\ \bottomrule
\end{tabular}}
\label{tab: train time}
\vspace{-2ex}
\end{table}

\vspace{-1ex}
\smallskip
\noindent\textbf{Remark. }Recall that simply injecting item embeddings or user representations from CF-SRec into LLMs, as done in existing LLM4Rec models, is insufficient for effective sequence understanding as shown in Sec.~\ref{sec: sequence exp}. In contrast, our ablation studies validate the effectiveness of our simple approach of incorporating the sequential knowledge into LLMs.

\subsection{Model analysis}

\subsubsection{\textbf{Train/Inference Efficiency.}}
\label{exp: time efficiency}
Note that \proposed~is efficient as it does not require fine-tuning either CF-SRec or the LLM itself. To quantitatively analyze the model efficiency, we compare the training and inference time of ~\proposed~ with LLM4Rec baselines (i.e., TALLRec, LLaRA, CoLLM, and A-LLMRec). Specifically, we measure the training time per epoch and the total inference time on the Scientific and Electronics datasets.
As shown in Table \ref{tab: train time}, \proposed~ achieves significantly faster training and inference times than all baselines. This is mainly because baselines using Parameter-Efficient Fine-Tuning methods such as LoRA require fine-tuning the LLMs, increasing both training and inference time. Furthermore, compared to A-LLMRec, which does not fine-tune the LLM, \proposed~ remains more efficient. This is mainly because A-LLMRec involves two-stage learning, leading to increase training time. Additionally, since A-LLMRec incorporates user representations into its prompts, it requires longer prompts during inference, resulting in higher computational overhead and slower inference speed compared to \proposed. These results highlight the efficiency of \proposed, supporting the model as a more computationally feasible solution for real-world applications for recommendation tasks while maintaining superior recommendation performance.

\subsubsection{\textbf{Size of LLMs}}
\label{exp: LLM size}
Note that all baseline LLM4Rec models including~\proposed~use LLaMA 3.2 (3B-Instruct) as the backbone LLMs. In this section, to investigate the impact of LLM size on recommendation performance, we replace the backbone with larger LLMs, i.e., LLaMA 3.1 (8B) \cite{llama3modelcard}.
We have the following observations in Figure~\ref{fig: llm size}:
1) Replacing the backbone LLMs to larger LLMs greatly enhances the overall performance of LLM4Rec models.
This aligns well with the scaling law of LLMs observed in other domains \cite{kaplan2020scaling}. Moreover, the superior performance of~\proposed~is still valid when the backbone is replaced to larger LLMs. 
2) Surprisingly, \proposed~with the smaller backbone LLMs outperforms the baseline LLM4Rec models with the larger backbone LLMs. This indicates that distilling sequential information is more crucial than merely increasing the LLM size to enhance the overall performance of sequential recommendation, which again highlights the importance of equipping LLMs with sequential knowledge to improve the sequential recommendation capabilities of LLMs.

\begin{figure}[t]
    \centering
    \includegraphics[width=0.99\linewidth]{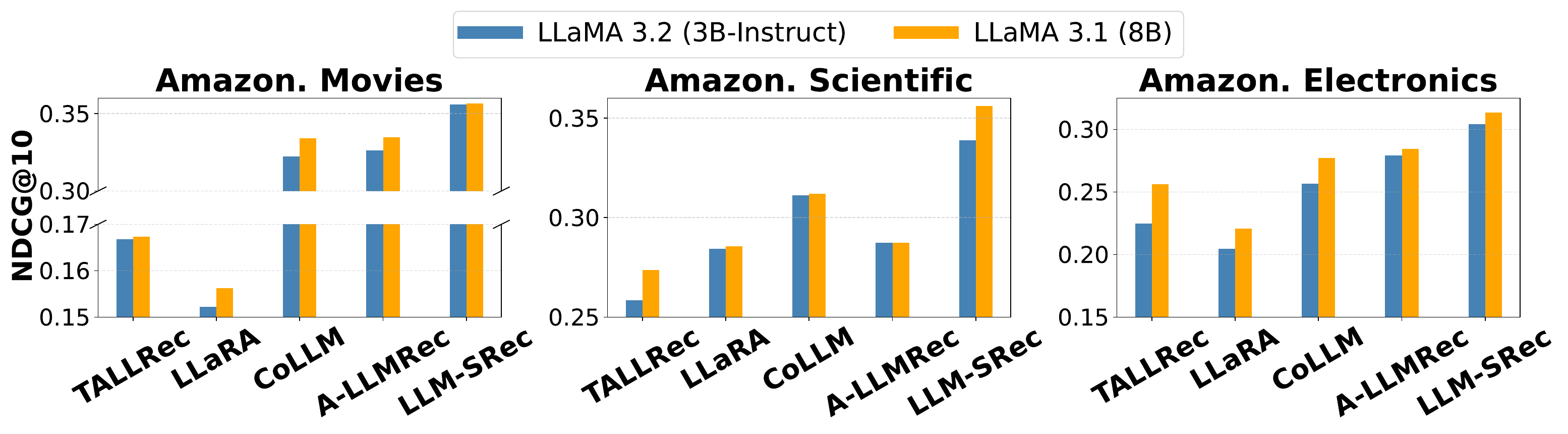}
    \vspace{-2ex}
    \caption
    {Performance of different sizes of backbone LLMs.}
    \label{fig: llm size}
    \vspace{-2.5ex}
\end{figure}

\vspace{-2ex}
\subsubsection{\textbf{Case Study.}}
\label{sec: case study}
In this section, we conduct a case study on the Electronics dataset to qualitatively validate the benefit of sequential knowledge and LLMs' textual understanding.
Figure \ref{fig: case study} shows three cases highlighting the effect of sequential knowledge and textual knowledge on recommendation. The cases are categorized as follows: (a) only~\proposed~provides correct recommendations, (b) LLM4Rec models (i.e., A-LLMRec and~\proposed) provide correct recommendations while SASRec fails, and (c)~\proposed~and SASRec provide correct recommendations while an LLM4Rec (i.e., A-LLMRec) baseline fails.
We have the following observations: 1) In case (a), the user's preference shifts from cable-related items to products from "Apple" brand. \proposed~correctly recommends  "Apple Pencil," while SASRec captures the changing preference but fails to recognize the textual information "Apple," leading to a wrong recommendation of an "Amazon Fire 7 Tablet." On the other hand, A-LLMRec, which struggles to capture sequential information, recommends "Audio Cable" based on textual knowledge of "Cable" and "Speaker." This emphasizes the importance of leveraging both sequential and textual information.
2) In case (b), the user focused on "BOSE" brand products. Both \proposed~and A-LLMRec, leveraging textual knowledge, successfully recommended the "BOSE" earbuds, while SASRec, lacking textual information, recommended the "SAMSUNG Galaxy Buds." This highlights the importance of textual knowledge in generating accurate recommendations.
3) In case (c), the user's preference shifts from "Hosa Tech" cables to security camera. \proposed~and SASRec capture this preference shift and provide relevant recommendations, while A-LLMRec recommends another "Hosa Tech" cable ignoring the sequential patterns.
Those cases demonstrate that both textual and sequential information are crucial for accurate recommendations, showcasing the superiority of \proposed~in integrating both for improved performance. 

\begin{figure}[t]
    \centering
    \includegraphics[width=0.95\linewidth]{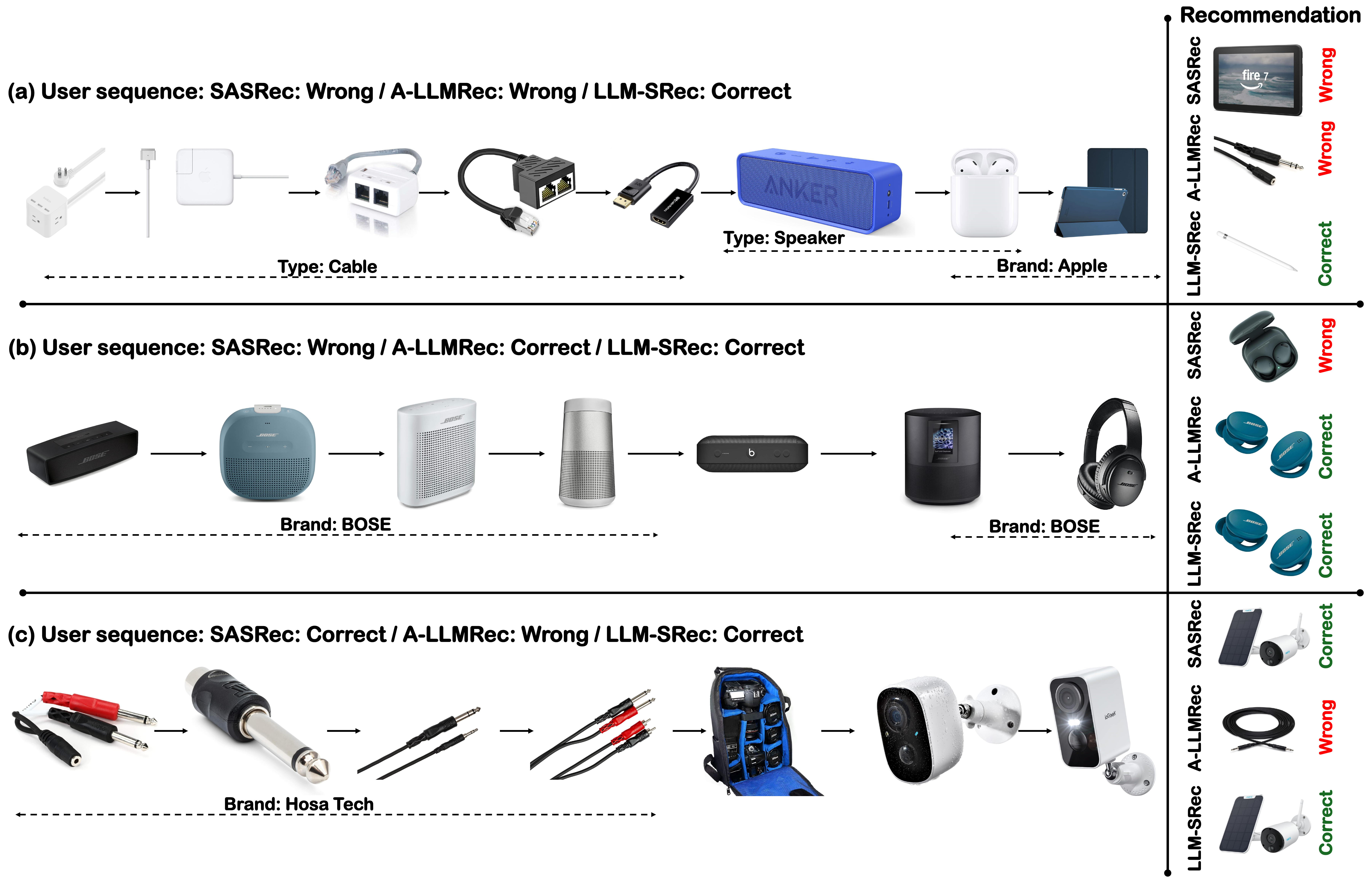}
    \vspace{-2ex}
    \caption
    {Case study on Electronics dataset.}
    \label{fig: case study}
    \vspace{-3.5ex}
\end{figure}

\vspace{-1ex}
\section{Related Work}
\noindent\textbf{Sequential Recommender Systems. }
Recommendation systems primarily focus on capturing collaborative filtering (CF) to identify similar items/users. Matrix Factorization-based approaches, achieved notable success by constructing CF knowledge in a latent space \cite{mnih2007probabilistic, chaney2015probabilistic, he2017neural, kim2025disentangling}. However, in conjunction with CF knowledge, understanding dynamic evolution in temporal user preferences has become a powerful tool, leading to the development of collaborative filtering-based sequential recommenders (CF-SRec) \cite{kang2018self, sun2019bert4rec, wu2019session, kim2023task, hidasi2015session,oh2023muse, choi2025dynamic}.
Initial approaches combined Matrix Factorization with Markov Chains to model temporal dynamics \cite{rendle2010factorizing}. Subsequently, neural network-based methods advanced sequential recommender systems, with GRU4Rec \cite{hidasi2015session} leveraging recurrent architectures, while methods such as Caser \cite{tang2018personalized} and NextItNet \cite{yuan2019simple} adopted Convolutional Neural Networks \cite{krizhevsky2012imagenet}. More recently, models such as SASRec \cite{yuan2019simple} and BERT4Rec \cite{sun2019bert4rec}, based on attention mechanisms, have demonstrated superior performance by focusing on the more relevant interaction sequences. These advancements underscore the importance of effectively modeling user behavior dynamics for improved recommendation accuracy.

\smallskip
\noindent\textbf{LLM-based Recommender Systems. }
LLMs have recently gained attention in recommendation systems \cite{yue2023llamarec, harte2023leveraging, dai2023uncovering, wu2024coral}, leveraging their reasoning ability and textual understanding for novel approaches such as zero-shot recommendation \cite{hou2024large} and conversational recommendation \cite{sanner2023large}. However, TALLRec \cite{bao2023tallrec} highlights the gap between LLMs' language modeling tasks and recommendation tasks, proposing a fine-tuning approach through Parameter-Efficient Fine-Tuning (PEFT) to adapt LLMs to recommendation tasks.
More recently, LLaRA \cite{10.1145/3626772.3657690}, CoLLM \cite{zhang2023collm}, and A-LLMRec \cite{10.1145/3637528.3671931} have been proposed. LLaRA and CoLLM combine CF-SRec item embeddings with text embeddings from item titles, enabling LLMs to utilize CF knowledge. A-LLMRec further incorporates item descriptions into a latent space, enabling the model to demonstrate robust performance in various scenarios.
Despite these advancements, prior methods fail to capture dynamic user preferences inherent in user interaction sequences as shown in Sec.~\ref{sec: sequence exp}. 

\vspace{-1.2ex}
\section{Conclusion}
In this paper, we address a fundamental limitation of LLM4Rec models, i.e., their inability to capture sequential patterns, and empirically demonstrate this shortcoming through extensive experiments. To address the limitation, we propose a simple yet effective distillation framework, named ~\proposed, which effectively transfers sequential knowledge extracted from CF-SRec into LLMs. By doing so, ~\proposed~ enables LLMs to effectively capture sequential dependencies, leading to superior recommendation performance compared to existing CF-SRec, LM-based recommender systems, and LLM4Rec. Furthermore, ~\proposed~ achieves high efficiency, as it does not require fine-tuning either CF-SRec or the LLM, demonstrating the effectiveness of our simple yet efficient architecture.

\vspace{-1ex}
\begin{acks}
This work was supported by NAVER Corporation, the National Research Foundation of Korea(NRF) grant funded by the Korea government(MSIT) (RS-2024-00335098), and the National Research Foundation of Korea(NRF) funded by Ministry of Science and ICT (RS-2022-NR068758).
\end{acks}

\clearpage
\bibliographystyle{ACM-Reference-Format}
\bibliography{sample-base}


\begin{thebibliography}{42}


\ifx \showCODEN    \undefined \def \showCODEN     #1{\unskip}     \fi
\ifx \showISBNx    \undefined \def \showISBNx     #1{\unskip}     \fi
\ifx \showISBNxiii \undefined \def \showISBNxiii  #1{\unskip}     \fi
\ifx \showISSN     \undefined \def \showISSN      #1{\unskip}     \fi
\ifx \showLCCN     \undefined \def \showLCCN      #1{\unskip}     \fi
\ifx \shownote     \undefined \def \shownote      #1{#1}          \fi
\ifx \showarticletitle \undefined \def \showarticletitle #1{#1}   \fi
\ifx \showURL      \undefined \def \showURL       {\relax}        \fi
\providecommand\bibfield[2]{#2}
\providecommand\bibinfo[2]{#2}
\providecommand\natexlab[1]{#1}
\providecommand\showeprint[2][]{arXiv:#2}

\bibitem[AI@Meta(2024)]%
        {llama3modelcard}
\bibfield{author}{\bibinfo{person}{AI@Meta}.} \bibinfo{year}{2024}\natexlab{}.
\newblock \showarticletitle{Llama 3 Model Card}.
\newblock  (\bibinfo{year}{2024}).
\newblock
\urldef\tempurl%
\url{https://github.com/meta-llama/llama3/blob/main/MODEL_CARD.md}
\showURL{%
\tempurl}


\bibitem[Bao et~al\mbox{.}(2023)]%
        {bao2023tallrec}
\bibfield{author}{\bibinfo{person}{Keqin Bao}, \bibinfo{person}{Jizhi Zhang}, \bibinfo{person}{Yang Zhang}, \bibinfo{person}{Wenjie Wang}, \bibinfo{person}{Fuli Feng}, {and} \bibinfo{person}{Xiangnan He}.} \bibinfo{year}{2023}\natexlab{}.
\newblock \showarticletitle{Tallrec: An effective and efficient tuning framework to align large language model with recommendation}. In \bibinfo{booktitle}{\emph{Proceedings of the 17th ACM Conference on Recommender Systems}}. \bibinfo{pages}{1007--1014}.
\newblock


\bibitem[Chaney et~al\mbox{.}(2015)]%
        {chaney2015probabilistic}
\bibfield{author}{\bibinfo{person}{Allison~JB Chaney}, \bibinfo{person}{David~M Blei}, {and} \bibinfo{person}{Tina Eliassi-Rad}.} \bibinfo{year}{2015}\natexlab{}.
\newblock \showarticletitle{A probabilistic model for using social networks in personalized item recommendation}. In \bibinfo{booktitle}{\emph{Proceedings of the 9th ACM Conference on Recommender Systems}}. \bibinfo{pages}{43--50}.
\newblock


\bibitem[Choi et~al\mbox{.}(2025)]%
        {choi2025dynamic}
\bibfield{author}{\bibinfo{person}{Seungyoon Choi}, \bibinfo{person}{Sein Kim}, \bibinfo{person}{Hongseok Kang}, \bibinfo{person}{Wonjoong Kim}, {and} \bibinfo{person}{Chanyoung Park}.} \bibinfo{year}{2025}\natexlab{}.
\newblock \showarticletitle{Dynamic Time-aware Continual User Representation Learning}.
\newblock \bibinfo{journal}{\emph{arXiv preprint arXiv:2504.16501}} (\bibinfo{year}{2025}).
\newblock


\bibitem[Dai et~al\mbox{.}(2023)]%
        {dai2023uncovering}
\bibfield{author}{\bibinfo{person}{Sunhao Dai}, \bibinfo{person}{Ninglu Shao}, \bibinfo{person}{Haiyuan Zhao}, \bibinfo{person}{Weijie Yu}, \bibinfo{person}{Zihua Si}, \bibinfo{person}{Chen Xu}, \bibinfo{person}{Zhongxiang Sun}, \bibinfo{person}{Xiao Zhang}, {and} \bibinfo{person}{Jun Xu}.} \bibinfo{year}{2023}\natexlab{}.
\newblock \showarticletitle{Uncovering chatgpt’s capabilities in recommender systems}. In \bibinfo{booktitle}{\emph{Proceedings of the 17th ACM Conference on Recommender Systems}}. \bibinfo{pages}{1126--1132}.
\newblock


\bibitem[Geng et~al\mbox{.}(2022)]%
        {geng2022recommendation}
\bibfield{author}{\bibinfo{person}{Shijie Geng}, \bibinfo{person}{Shuchang Liu}, \bibinfo{person}{Zuohui Fu}, \bibinfo{person}{Yingqiang Ge}, {and} \bibinfo{person}{Yongfeng Zhang}.} \bibinfo{year}{2022}\natexlab{}.
\newblock \showarticletitle{Recommendation as language processing (rlp): A unified pretrain, personalized prompt \& predict paradigm (p5)}. In \bibinfo{booktitle}{\emph{Proceedings of the 16th ACM Conference on Recommender Systems}}. \bibinfo{pages}{299--315}.
\newblock


\bibitem[Harte et~al\mbox{.}(2023)]%
        {harte2023leveraging}
\bibfield{author}{\bibinfo{person}{Jesse Harte}, \bibinfo{person}{Wouter Zorgdrager}, \bibinfo{person}{Panos Louridas}, \bibinfo{person}{Asterios Katsifodimos}, \bibinfo{person}{Dietmar Jannach}, {and} \bibinfo{person}{Marios Fragkoulis}.} \bibinfo{year}{2023}\natexlab{}.
\newblock \showarticletitle{Leveraging large language models for sequential recommendation}. In \bibinfo{booktitle}{\emph{Proceedings of the 17th ACM Conference on Recommender Systems}}. \bibinfo{pages}{1096--1102}.
\newblock


\bibitem[He et~al\mbox{.}(2017)]%
        {he2017neural}
\bibfield{author}{\bibinfo{person}{Xiangnan He}, \bibinfo{person}{Lizi Liao}, \bibinfo{person}{Hanwang Zhang}, \bibinfo{person}{Liqiang Nie}, \bibinfo{person}{Xia Hu}, {and} \bibinfo{person}{Tat-Seng Chua}.} \bibinfo{year}{2017}\natexlab{}.
\newblock \showarticletitle{Neural collaborative filtering}. In \bibinfo{booktitle}{\emph{Proceedings of the 26th international conference on world wide web}}. \bibinfo{pages}{173--182}.
\newblock


\bibitem[Hidasi et~al\mbox{.}(2015)]%
        {hidasi2015session}
\bibfield{author}{\bibinfo{person}{Bal{\'a}zs Hidasi}, \bibinfo{person}{Alexandros Karatzoglou}, \bibinfo{person}{Linas Baltrunas}, {and} \bibinfo{person}{Domonkos Tikk}.} \bibinfo{year}{2015}\natexlab{}.
\newblock \showarticletitle{Session-based recommendations with recurrent neural networks}.
\newblock \bibinfo{journal}{\emph{arXiv preprint arXiv:1511.06939}} (\bibinfo{year}{2015}).
\newblock


\bibitem[Hou et~al\mbox{.}(2024a)]%
        {hou2024bridging}
\bibfield{author}{\bibinfo{person}{Yupeng Hou}, \bibinfo{person}{Jiacheng Li}, \bibinfo{person}{Zhankui He}, \bibinfo{person}{An Yan}, \bibinfo{person}{Xiusi Chen}, {and} \bibinfo{person}{Julian McAuley}.} \bibinfo{year}{2024}\natexlab{a}.
\newblock \showarticletitle{Bridging Language and Items for Retrieval and Recommendation}.
\newblock \bibinfo{journal}{\emph{arXiv preprint arXiv:2403.03952}} (\bibinfo{year}{2024}).
\newblock


\bibitem[Hou et~al\mbox{.}(2024b)]%
        {hou2024large}
\bibfield{author}{\bibinfo{person}{Yupeng Hou}, \bibinfo{person}{Junjie Zhang}, \bibinfo{person}{Zihan Lin}, \bibinfo{person}{Hongyu Lu}, \bibinfo{person}{Ruobing Xie}, \bibinfo{person}{Julian McAuley}, {and} \bibinfo{person}{Wayne~Xin Zhao}.} \bibinfo{year}{2024}\natexlab{b}.
\newblock \showarticletitle{Large language models are zero-shot rankers for recommender systems}. In \bibinfo{booktitle}{\emph{European Conference on Information Retrieval}}. Springer, \bibinfo{pages}{364--381}.
\newblock


\bibitem[Hu et~al\mbox{.}(2022)]%
        {hu2022lora}
\bibfield{author}{\bibinfo{person}{Edward~J Hu}, \bibinfo{person}{yelong shen}, \bibinfo{person}{Phillip Wallis}, \bibinfo{person}{Zeyuan Allen-Zhu}, \bibinfo{person}{Yuanzhi Li}, \bibinfo{person}{Shean Wang}, \bibinfo{person}{Lu Wang}, {and} \bibinfo{person}{Weizhu Chen}.} \bibinfo{year}{2022}\natexlab{}.
\newblock \showarticletitle{Lo{RA}: Low-Rank Adaptation of Large Language Models}. In \bibinfo{booktitle}{\emph{International Conference on Learning Representations}}.
\newblock


\bibitem[Kang and McAuley(2018)]%
        {kang2018self}
\bibfield{author}{\bibinfo{person}{Wang-Cheng Kang} {and} \bibinfo{person}{Julian McAuley}.} \bibinfo{year}{2018}\natexlab{}.
\newblock \showarticletitle{Self-attentive sequential recommendation}. In \bibinfo{booktitle}{\emph{2018 IEEE international conference on data mining (ICDM)}}. IEEE, \bibinfo{pages}{197--206}.
\newblock


\bibitem[Kaplan et~al\mbox{.}(2020)]%
        {kaplan2020scaling}
\bibfield{author}{\bibinfo{person}{Jared Kaplan}, \bibinfo{person}{Sam McCandlish}, \bibinfo{person}{Tom Henighan}, \bibinfo{person}{Tom~B Brown}, \bibinfo{person}{Benjamin Chess}, \bibinfo{person}{Rewon Child}, \bibinfo{person}{Scott Gray}, \bibinfo{person}{Alec Radford}, \bibinfo{person}{Jeffrey Wu}, {and} \bibinfo{person}{Dario Amodei}.} \bibinfo{year}{2020}\natexlab{}.
\newblock \showarticletitle{Scaling laws for neural language models}.
\newblock \bibinfo{journal}{\emph{arXiv preprint arXiv:2001.08361}} (\bibinfo{year}{2020}).
\newblock


\bibitem[Kim et~al\mbox{.}(2025)]%
        {kim2025disentangling}
\bibfield{author}{\bibinfo{person}{Jiwan Kim}, \bibinfo{person}{Hongseok Kang}, \bibinfo{person}{Sein Kim}, \bibinfo{person}{Kibum Kim}, {and} \bibinfo{person}{Chanyoung Park}.} \bibinfo{year}{2025}\natexlab{}.
\newblock \showarticletitle{Disentangling and Generating Modalities for Recommendation in Missing Modality Scenarios}.
\newblock \bibinfo{journal}{\emph{arXiv preprint arXiv:2504.16352}} (\bibinfo{year}{2025}).
\newblock


\bibitem[Kim et~al\mbox{.}(2024)]%
        {10.1145/3637528.3671931}
\bibfield{author}{\bibinfo{person}{Sein Kim}, \bibinfo{person}{Hongseok Kang}, \bibinfo{person}{Seungyoon Choi}, \bibinfo{person}{Donghyun Kim}, \bibinfo{person}{Minchul Yang}, {and} \bibinfo{person}{Chanyoung Park}.} \bibinfo{year}{2024}\natexlab{}.
\newblock \showarticletitle{Large Language Models meet Collaborative Filtering: An Efficient All-round LLM-based Recommender System}. In \bibinfo{booktitle}{\emph{Proceedings of the 30th ACM SIGKDD Conference on Knowledge Discovery and Data Mining}} (Barcelona, Spain) \emph{(\bibinfo{series}{KDD '24})}. \bibinfo{publisher}{Association for Computing Machinery}, \bibinfo{address}{New York, NY, USA}, \bibinfo{pages}{1395–1406}.
\newblock
\showISBNx{9798400704901}
\href{https://doi.org/10.1145/3637528.3671931}{doi:\nolinkurl{10.1145/3637528.3671931}}


\bibitem[Kim et~al\mbox{.}(2023)]%
        {kim2023task}
\bibfield{author}{\bibinfo{person}{Sein Kim}, \bibinfo{person}{Namkyeong Lee}, \bibinfo{person}{Donghyun Kim}, \bibinfo{person}{Minchul Yang}, {and} \bibinfo{person}{Chanyoung Park}.} \bibinfo{year}{2023}\natexlab{}.
\newblock \showarticletitle{Task Relation-aware Continual User Representation Learning}. In \bibinfo{booktitle}{\emph{Proceedings of the 29th ACM SIGKDD Conference on Knowledge Discovery and Data Mining}}. \bibinfo{pages}{1107--1119}.
\newblock


\bibitem[Klenitskiy et~al\mbox{.}(2024)]%
        {10.1145/3640457.3688195}
\bibfield{author}{\bibinfo{person}{Anton Klenitskiy}, \bibinfo{person}{Anna Volodkevich}, \bibinfo{person}{Anton Pembek}, {and} \bibinfo{person}{Alexey Vasilev}.} \bibinfo{year}{2024}\natexlab{}.
\newblock \showarticletitle{Does It Look Sequential? An Analysis of Datasets for Evaluation of Sequential Recommendations}. In \bibinfo{booktitle}{\emph{Proceedings of the 18th ACM Conference on Recommender Systems}} (Bari, Italy) \emph{(\bibinfo{series}{RecSys '24})}. \bibinfo{publisher}{Association for Computing Machinery}, \bibinfo{address}{New York, NY, USA}, \bibinfo{pages}{1067–1072}.
\newblock
\showISBNx{9798400705052}
\href{https://doi.org/10.1145/3640457.3688195}{doi:\nolinkurl{10.1145/3640457.3688195}}


\bibitem[Krizhevsky et~al\mbox{.}(2012)]%
        {krizhevsky2012imagenet}
\bibfield{author}{\bibinfo{person}{Alex Krizhevsky}, \bibinfo{person}{Ilya Sutskever}, {and} \bibinfo{person}{Geoffrey~E Hinton}.} \bibinfo{year}{2012}\natexlab{}.
\newblock \showarticletitle{Imagenet classification with deep convolutional neural networks}.
\newblock \bibinfo{journal}{\emph{Advances in neural information processing systems}}  \bibinfo{volume}{25} (\bibinfo{year}{2012}).
\newblock


\bibitem[Li et~al\mbox{.}(2023c)]%
        {li2023text}
\bibfield{author}{\bibinfo{person}{Jiacheng Li}, \bibinfo{person}{Ming Wang}, \bibinfo{person}{Jin Li}, \bibinfo{person}{Jinmiao Fu}, \bibinfo{person}{Xin Shen}, \bibinfo{person}{Jingbo Shang}, {and} \bibinfo{person}{Julian McAuley}.} \bibinfo{year}{2023}\natexlab{c}.
\newblock \showarticletitle{Text is all you need: Learning language representations for sequential recommendation}. In \bibinfo{booktitle}{\emph{Proceedings of the 29th ACM SIGKDD Conference on Knowledge Discovery and Data Mining}}. \bibinfo{pages}{1258--1267}.
\newblock


\bibitem[Li et~al\mbox{.}(2023a)]%
        {li2023ctrl}
\bibfield{author}{\bibinfo{person}{Xiangyang Li}, \bibinfo{person}{Bo Chen}, \bibinfo{person}{Lu Hou}, {and} \bibinfo{person}{Ruiming Tang}.} \bibinfo{year}{2023}\natexlab{a}.
\newblock \showarticletitle{Ctrl: Connect tabular and language model for ctr prediction}.
\newblock \bibinfo{journal}{\emph{CoRR}} (\bibinfo{year}{2023}).
\newblock


\bibitem[Li et~al\mbox{.}(2023b)]%
        {li2023e4srec}
\bibfield{author}{\bibinfo{person}{Xinhang Li}, \bibinfo{person}{Chong Chen}, \bibinfo{person}{Xiangyu Zhao}, \bibinfo{person}{Yong Zhang}, {and} \bibinfo{person}{Chunxiao Xing}.} \bibinfo{year}{2023}\natexlab{b}.
\newblock \showarticletitle{E4srec: An elegant effective efficient extensible solution of large language models for sequential recommendation}.
\newblock \bibinfo{journal}{\emph{arXiv preprint arXiv:2312.02443}} (\bibinfo{year}{2023}).
\newblock


\bibitem[Liao et~al\mbox{.}(2024)]%
        {10.1145/3626772.3657690}
\bibfield{author}{\bibinfo{person}{Jiayi Liao}, \bibinfo{person}{Sihang Li}, \bibinfo{person}{Zhengyi Yang}, \bibinfo{person}{Jiancan Wu}, \bibinfo{person}{Yancheng Yuan}, \bibinfo{person}{Xiang Wang}, {and} \bibinfo{person}{Xiangnan He}.} \bibinfo{year}{2024}\natexlab{}.
\newblock \showarticletitle{LLaRA: Large Language-Recommendation Assistant}. In \bibinfo{booktitle}{\emph{Proceedings of the 47th International ACM SIGIR Conference on Research and Development in Information Retrieval}} (Washington DC, USA) \emph{(\bibinfo{series}{SIGIR '24})}. \bibinfo{publisher}{Association for Computing Machinery}, \bibinfo{address}{New York, NY, USA}, \bibinfo{pages}{1785–1795}.
\newblock
\showISBNx{9798400704314}
\href{https://doi.org/10.1145/3626772.3657690}{doi:\nolinkurl{10.1145/3626772.3657690}}


\bibitem[Liu and Abbeel(2024)]%
        {liu2024blockwise}
\bibfield{author}{\bibinfo{person}{Hao Liu} {and} \bibinfo{person}{Pieter Abbeel}.} \bibinfo{year}{2024}\natexlab{}.
\newblock \showarticletitle{Blockwise parallel transformers for large context models}.
\newblock \bibinfo{journal}{\emph{Advances in Neural Information Processing Systems}}  \bibinfo{volume}{36} (\bibinfo{year}{2024}).
\newblock


\bibitem[Liu(2019)]%
        {liu2019roberta}
\bibfield{author}{\bibinfo{person}{Yinhan Liu}.} \bibinfo{year}{2019}\natexlab{}.
\newblock \showarticletitle{Roberta: A robustly optimized bert pretraining approach}.
\newblock \bibinfo{journal}{\emph{arXiv preprint arXiv:1907.11692}}  \bibinfo{volume}{364} (\bibinfo{year}{2019}).
\newblock


\bibitem[Mnih and Salakhutdinov(2007)]%
        {mnih2007probabilistic}
\bibfield{author}{\bibinfo{person}{Andriy Mnih} {and} \bibinfo{person}{Russ~R Salakhutdinov}.} \bibinfo{year}{2007}\natexlab{}.
\newblock \showarticletitle{Probabilistic matrix factorization}.
\newblock \bibinfo{journal}{\emph{Advances in neural information processing systems}}  \bibinfo{volume}{20} (\bibinfo{year}{2007}).
\newblock


\bibitem[Oh et~al\mbox{.}(2023)]%
        {oh2023muse}
\bibfield{author}{\bibinfo{person}{Yunhak Oh}, \bibinfo{person}{Sukwon Yun}, \bibinfo{person}{Dongmin Hyun}, \bibinfo{person}{Sein Kim}, {and} \bibinfo{person}{Chanyoung Park}.} \bibinfo{year}{2023}\natexlab{}.
\newblock \showarticletitle{Muse: music recommender system with shuffle play recommendation enhancement}. In \bibinfo{booktitle}{\emph{Proceedings of the 32nd ACM international conference on information and knowledge management}}. \bibinfo{pages}{1928--1938}.
\newblock


\bibitem[Rendle et~al\mbox{.}(2010)]%
        {rendle2010factorizing}
\bibfield{author}{\bibinfo{person}{Steffen Rendle}, \bibinfo{person}{Christoph Freudenthaler}, {and} \bibinfo{person}{Lars Schmidt-Thieme}.} \bibinfo{year}{2010}\natexlab{}.
\newblock \showarticletitle{Factorizing personalized markov chains for next-basket recommendation}. In \bibinfo{booktitle}{\emph{Proceedings of the 19th international conference on World wide web}}. \bibinfo{pages}{811--820}.
\newblock


\bibitem[Sanner et~al\mbox{.}(2023)]%
        {sanner2023large}
\bibfield{author}{\bibinfo{person}{Scott Sanner}, \bibinfo{person}{Krisztian Balog}, \bibinfo{person}{Filip Radlinski}, \bibinfo{person}{Ben Wedin}, {and} \bibinfo{person}{Lucas Dixon}.} \bibinfo{year}{2023}\natexlab{}.
\newblock \showarticletitle{Large language models are competitive near cold-start recommenders for language-and item-based preferences}. In \bibinfo{booktitle}{\emph{Proceedings of the 17th ACM conference on recommender systems}}. \bibinfo{pages}{890--896}.
\newblock


\bibitem[Sun et~al\mbox{.}(2019)]%
        {sun2019bert4rec}
\bibfield{author}{\bibinfo{person}{Fei Sun}, \bibinfo{person}{Jun Liu}, \bibinfo{person}{Jian Wu}, \bibinfo{person}{Changhua Pei}, \bibinfo{person}{Xiao Lin}, \bibinfo{person}{Wenwu Ou}, {and} \bibinfo{person}{Peng Jiang}.} \bibinfo{year}{2019}\natexlab{}.
\newblock \showarticletitle{BERT4Rec: Sequential recommendation with bidirectional encoder representations from transformer}. In \bibinfo{booktitle}{\emph{Proceedings of the 28th ACM international conference on information and knowledge management}}. \bibinfo{pages}{1441--1450}.
\newblock


\bibitem[Takahagi and Shinnou(2023)]%
        {takahagi-shinnou-2023-data}
\bibfield{author}{\bibinfo{person}{Kyosuke Takahagi} {and} \bibinfo{person}{Hiroyuki Shinnou}.} \bibinfo{year}{2023}\natexlab{}.
\newblock \showarticletitle{Data Augmentation by Shuffling Phrases in Recognizing Textual Entailment}. In \bibinfo{booktitle}{\emph{Proceedings of the 37th Pacific Asia Conference on Language, Information and Computation}}. \bibinfo{publisher}{Association for Computational Linguistics}, \bibinfo{address}{Hong Kong, China}, \bibinfo{pages}{194--200}.
\newblock
\urldef\tempurl%
\url{https://aclanthology.org/2023.paclic-1.19/}
\showURL{%
\tempurl}


\bibitem[Tang and Wang(2018a)]%
        {10.1145/3159652.3159656}
\bibfield{author}{\bibinfo{person}{Jiaxi Tang} {and} \bibinfo{person}{Ke Wang}.} \bibinfo{year}{2018}\natexlab{a}.
\newblock \showarticletitle{Personalized Top-N Sequential Recommendation via Convolutional Sequence Embedding}. In \bibinfo{booktitle}{\emph{Proceedings of the Eleventh ACM International Conference on Web Search and Data Mining}} (Marina Del Rey, CA, USA) \emph{(\bibinfo{series}{WSDM '18})}. \bibinfo{publisher}{Association for Computing Machinery}, \bibinfo{address}{New York, NY, USA}, \bibinfo{pages}{565–573}.
\newblock
\showISBNx{9781450355810}
\urldef\tempurl%
\url{https://doi.org/10.1145/3159652.3159656}
\showURL{%
\tempurl}


\bibitem[Tang and Wang(2018b)]%
        {tang2018personalized}
\bibfield{author}{\bibinfo{person}{Jiaxi Tang} {and} \bibinfo{person}{Ke Wang}.} \bibinfo{year}{2018}\natexlab{b}.
\newblock \showarticletitle{Personalized top-n sequential recommendation via convolutional sequence embedding}. In \bibinfo{booktitle}{\emph{Proceedings of the eleventh ACM international conference on web search and data mining}}. \bibinfo{pages}{565--573}.
\newblock


\bibitem[Wang et~al\mbox{.}(2024)]%
        {10.1145/3627673.3679535}
\bibfield{author}{\bibinfo{person}{Chen Wang}, \bibinfo{person}{Liangwei Yang}, \bibinfo{person}{Zhiwei Liu}, \bibinfo{person}{Xiaolong Liu}, \bibinfo{person}{Mingdai Yang}, \bibinfo{person}{Yueqing Liang}, {and} \bibinfo{person}{Philip~S. Yu}.} \bibinfo{year}{2024}\natexlab{}.
\newblock \showarticletitle{Collaborative Alignment for Recommendation}. In \bibinfo{booktitle}{\emph{Proceedings of the 33rd ACM International Conference on Information and Knowledge Management}} (Boise, ID, USA) \emph{(\bibinfo{series}{CIKM '24})}. \bibinfo{publisher}{Association for Computing Machinery}, \bibinfo{address}{New York, NY, USA}, \bibinfo{pages}{2315–2325}.
\newblock
\showISBNx{9798400704369}
\href{https://doi.org/10.1145/3627673.3679535}{doi:\nolinkurl{10.1145/3627673.3679535}}


\bibitem[Wang et~al\mbox{.}(2022)]%
        {10.1145/3534678.3539253}
\bibfield{author}{\bibinfo{person}{Chenyang Wang}, \bibinfo{person}{Yuanqing Yu}, \bibinfo{person}{Weizhi Ma}, \bibinfo{person}{Min Zhang}, \bibinfo{person}{Chong Chen}, \bibinfo{person}{Yiqun Liu}, {and} \bibinfo{person}{Shaoping Ma}.} \bibinfo{year}{2022}\natexlab{}.
\newblock \showarticletitle{Towards Representation Alignment and Uniformity in Collaborative Filtering}. In \bibinfo{booktitle}{\emph{Proceedings of the 28th ACM SIGKDD Conference on Knowledge Discovery and Data Mining}} (Washington DC, USA) \emph{(\bibinfo{series}{KDD '22})}. \bibinfo{publisher}{Association for Computing Machinery}, \bibinfo{address}{New York, NY, USA}, \bibinfo{pages}{1816–1825}.
\newblock
\showISBNx{9781450393850}
\href{https://doi.org/10.1145/3534678.3539253}{doi:\nolinkurl{10.1145/3534678.3539253}}


\bibitem[Wang and Isola(2020)]%
        {wang2020understanding}
\bibfield{author}{\bibinfo{person}{Tongzhou Wang} {and} \bibinfo{person}{Phillip Isola}.} \bibinfo{year}{2020}\natexlab{}.
\newblock \showarticletitle{Understanding contrastive representation learning through alignment and uniformity on the hypersphere}. In \bibinfo{booktitle}{\emph{International conference on machine learning}}. PMLR, \bibinfo{pages}{9929--9939}.
\newblock


\bibitem[Woolridge et~al\mbox{.}(2021)]%
        {woolridge2021sequence}
\bibfield{author}{\bibinfo{person}{Daniel Woolridge}, \bibinfo{person}{Sean Wilner}, {and} \bibinfo{person}{Madeleine Glick}.} \bibinfo{year}{2021}\natexlab{}.
\newblock \showarticletitle{Sequence or Pseudo-Sequence? An Analysis of Sequential Recommendation Datasets.}. In \bibinfo{booktitle}{\emph{Perspectives@ RecSys}}.
\newblock


\bibitem[Wu et~al\mbox{.}(2024)]%
        {wu2024coral}
\bibfield{author}{\bibinfo{person}{Junda Wu}, \bibinfo{person}{Cheng-Chun Chang}, \bibinfo{person}{Tong Yu}, \bibinfo{person}{Zhankui He}, \bibinfo{person}{Jianing Wang}, \bibinfo{person}{Yupeng Hou}, {and} \bibinfo{person}{Julian McAuley}.} \bibinfo{year}{2024}\natexlab{}.
\newblock \showarticletitle{Coral: Collaborative retrieval-augmented large language models improve long-tail recommendation}. In \bibinfo{booktitle}{\emph{Proceedings of the 30th ACM SIGKDD Conference on Knowledge Discovery and Data Mining}}. \bibinfo{pages}{3391--3401}.
\newblock


\bibitem[Wu et~al\mbox{.}(2019)]%
        {wu2019session}
\bibfield{author}{\bibinfo{person}{Shu Wu}, \bibinfo{person}{Yuyuan Tang}, \bibinfo{person}{Yanqiao Zhu}, \bibinfo{person}{Liang Wang}, \bibinfo{person}{Xing Xie}, {and} \bibinfo{person}{Tieniu Tan}.} \bibinfo{year}{2019}\natexlab{}.
\newblock \showarticletitle{Session-based recommendation with graph neural networks}. In \bibinfo{booktitle}{\emph{Proceedings of the AAAI conference on artificial intelligence}}, Vol.~\bibinfo{volume}{33}. \bibinfo{pages}{346--353}.
\newblock


\bibitem[Yuan et~al\mbox{.}(2019)]%
        {yuan2019simple}
\bibfield{author}{\bibinfo{person}{Fajie Yuan}, \bibinfo{person}{Alexandros Karatzoglou}, \bibinfo{person}{Ioannis Arapakis}, \bibinfo{person}{Joemon~M Jose}, {and} \bibinfo{person}{Xiangnan He}.} \bibinfo{year}{2019}\natexlab{}.
\newblock \showarticletitle{A simple convolutional generative network for next item recommendation}. In \bibinfo{booktitle}{\emph{Proceedings of the twelfth ACM international conference on web search and data mining}}. \bibinfo{pages}{582--590}.
\newblock


\bibitem[Yue et~al\mbox{.}(2023)]%
        {yue2023llamarec}
\bibfield{author}{\bibinfo{person}{Zhenrui Yue}, \bibinfo{person}{Sara Rabhi}, \bibinfo{person}{Gabriel de Souza~Pereira Moreira}, \bibinfo{person}{Dong Wang}, {and} \bibinfo{person}{Even Oldridge}.} \bibinfo{year}{2023}\natexlab{}.
\newblock \showarticletitle{LlamaRec: Two-stage recommendation using large language models for ranking}.
\newblock \bibinfo{journal}{\emph{arXiv preprint arXiv:2311.02089}} (\bibinfo{year}{2023}).
\newblock


\bibitem[Zhang et~al\mbox{.}(2023)]%
        {zhang2023collm}
\bibfield{author}{\bibinfo{person}{Yang Zhang}, \bibinfo{person}{Fuli Feng}, \bibinfo{person}{Jizhi Zhang}, \bibinfo{person}{Keqin Bao}, \bibinfo{person}{Qifan Wang}, {and} \bibinfo{person}{Xiangnan He}.} \bibinfo{year}{2023}\natexlab{}.
\newblock \showarticletitle{Collm: Integrating collaborative embeddings into large language models for recommendation}.
\newblock \bibinfo{journal}{\emph{arXiv preprint arXiv:2310.19488}} (\bibinfo{year}{2023}).
\newblock


\end{thebibliography}

\appendix

\clearpage


\section{Details of LLM4Rec Prompt Construction}
This section provides additional details on how to construct prompts for LLM4Rec models discussed in Sec.~\ref{sec: problem setup}.
\subsection{Next Item Title Generation}
\label{app: next item title generation}
Based on the user interaction sequence $\mathcal{S}_u$ and candidate set $\mathcal{C}_u$ of user $u$, the textual data for the interacted items and candidate items are defined as $\mathcal{T}_{\mathcal{S}_u} = \left\{ \text{Text}(i) \mid i \in \mathcal{S}_u \right\}$ and $\mathcal{T}_{\mathcal{C}_u} = \left\{ \text{Text}(i) \mid i \in \mathcal{C}_u \right\}$, respectively, where $\text{Text}(i)$ represents textual information (e.g., title or description) of item $i$. 

For models such as LLaRA \cite{10.1145/3626772.3657690}, CoLLM \cite{zhang2023collm}, and A-LLMRec \cite{10.1145/3637528.3671931}, which incorporate item embeddings and user representations from a pre-trained CF-SRec, we use $\mathbf{E} \in \mathbb{R}^{|\mathcal{I}| \times d}$ to denote the item embedding matrix of the pre-trained CF-SRec, where $d$ is the hidden dimension of the embeddings. 
We also define $f_{\mathcal{I}}$ and $f_{\mathcal{U}}$ as the item and user projection layers used in LLaRA, CoLLM, and A-LLMRec (includes Stage-1 item encoder of A-LLMRec), respectively. 
The embeddings of items in the item interaction sequence $\mathcal{S}_u$ are defined as $\mathbf{E}_{\mathcal{S}_u} = \left\{ f_{\mathcal{I}}(\mathbf{E}_i) \mid i \in \mathcal{S}_u \right\}$, while the embeddings for the candidate items $\mathcal{C}_u$ are represented as $\mathbf{E}_{\mathcal{C}_u} = \left\{ f_{\mathcal{I}}(\mathbf{E}_i) \mid i \in \mathcal{C}_u \right\}${, where $f_{\mathcal{I}}(\mathbf{E_i})\in\mathbb{R}^{d_{llm}}$ and $d_{llm}$ denotes the token embedding dimension of LLM}. 
{Furthermore, the user representation is defined as $\mathbf{Z}_u = f_{\mathcal{U}}(\text{CF-SRec}(\mathcal{S}_u)) \in \mathbb{R}^{d_{llm}}$, where $\text{CF-SRec}(\mathcal{S}_u)$ represents the user $u$'s representation obtained from the item interaction sequence $\mathcal{S}_u$ using a pre-trained CF-SRec.} 
Then, using the prompts shown in Table~\ref{tab title generation prompt}, LLMs are trained for the sequential recommendation task through the Next Item Title Generation approach as follows:
\begin{equation}
    p(\text{Text}(i_{n_u+1}^{(u)}) \mid \mathcal{P}^{u}, \mathcal{D}^{u})
    \label{Eq LLM4Rec Title generation}
\end{equation}
where $\mathcal{P}^u$ is the input prompt for user $u$, and $D^u$ represents the set of interacted and candidate item titles and their corresponding embeddings used in Table \ref{tab title generation prompt} for user $u$ as follows:
\begin{align}
    \mathcal{D}^{u} = \begin{cases}
        \mathcal{T}_{\mathcal{S}_u}, \mathcal{T}_{\mathcal{C}_u} & \text{TALLRec}\\
        \mathcal{T}_{\mathcal{S}_u}, \mathcal{T}_{\mathcal{C}_u}, \mathbf{E}_{\mathcal{S}_u}, \mathbf{E}_{\mathcal{C}_u} & \text{LLaRA} \\
        \mathcal{T}_{\mathcal{S}_u}, \mathcal{T}_{\mathcal{C}_u},\mathbf{E}_{\mathcal{S}_u}, \mathbf{E}_{\mathcal{C}_u}, \mathbf{Z}_u & \text{CoLLM/A-LLMRec}
    \end{cases}
    \label{Eq LLM4Rec Title generation Input}
\end{align}

\subsection{Next Item Retrieval}
\label{app: next item retrieval}
Based on the prompt $\mathcal{P}^u_{\mathcal{U}}$ and $\mathcal{P}^i_{\mathcal{I}}$ in Table~\ref{tab next item retrieval prompt}, we extract representation of user $u\in\mathcal{U}$, denoted $\mathbf{h}^u_{\mathcal{U}}$ and the embedding of item $i\in\mathcal{C}_u$, denoted $\mathbf{h}^i_{\mathcal{I}}$ as follows:
\begin{align}
    \mathbf{h}^u_{\mathcal{U}} = \text{LLM}(\mathcal{P}^u_{\mathcal{U}}, \mathcal{D'}^u), \,\,\,\,
    \mathbf{h}^{i}_{\mathcal{I}} = \text{LLM}(\mathcal{P}^i_{\mathcal{I}},  \mathcal{D'}^i)
    \label{Eq LLM4Rec Retrieval}
\end{align}
where $\mathcal{P}^u_{\mathcal{U}}$ denotes the input prompt for user $u$ to extract representation of user $u$, $\mathcal{P}^i_{\mathcal{I}}$ denotes the input prompt for item $i$ to extract embedding of item $i$,
{$D'^u$ denotes the set of interacted item titles and their corresponding embeddings for user $u$, while $D'^i$ denotes the item title and its embedding for candidate item $i$, as presented in Table \ref{tab next item retrieval prompt}, as follows:}
\begin{align}
    \begin{split}
    \mathcal{D'}^{u} &= \begin{cases}
        \mathcal{T}_{\mathcal{S}_u} & \text{TALLRec}\\
        \mathcal{T}_{\mathcal{S}_u}, \mathbf{E}_{\mathcal{S}_u} & \text{LLaRA} \\
        \mathcal{T}_{\mathcal{S}_u},\mathbf{E}_{\mathcal{S}_u}, \mathbf{Z}_u & \text{CoLLM/A-LLMRec/\proposed}
    \end{cases}\\
    \mathcal{D'}^{i} &= \begin{cases}
        \text{Text}(i) & \text{TALLRec}\\
        \text{Text}(i), f_{\mathcal{I}}(\mathbf{E}_i) & \text{LLaRA/CoLLM/A-LLMRec/\proposed}
    \end{cases}
    \end{split}
    \label{Eq LLM4Rec Retrieval Input}
\end{align}

Then, using the user representation $\mathbf{h}^u_{\mathcal{U}}$ and item embedding $\mathbf{h}^i_{\mathcal{I}}$, LLMs are trained for the sequential recommendation task through the Next Item Retrieval approach as follows:
\begin{equation}
\small
p(i_{n_u+1}^{(u)} \mid \mathcal{S}_u) \propto s(u,i_{n_u+1}^{(u)}) = f_\mathit{item}(\mathbf{h}^{i_{n_u+1}^{(u)}}_{\mathcal{I}}) \cdot f_\mathit{user}(\mathbf{h}^u_{\mathcal{U}})^T
    \label{Eq next item retrieval probability}
\end{equation}
where $f_\mathit{item}$ and $f_\mathit{user}$ denote projection layers defined in Equation~\ref{Eq distill}.

\section{Datasets}
\label{app: dataset}
Table \ref{tab dataset} shows the statistics of the dataset after preprocessing.

\begin{table}[h]
\caption{Statistics of datasets after preprocessing.}
\vspace{-2ex}
\resizebox{0.75\linewidth}{!}{
\begin{tabular}{c|c|c|c|c}
\toprule
Dataset & Movies & Scientific & Electronics & CDs  \\ \midrule\midrule
\# Users      & 11,947    & 23,627    &  27,601 & 18,481 \\ \midrule
\# Items  & 17,490    & 25,764    &  31,533 & 30,951   \\ \midrule
\# Interactions & 144,071    & 266,164    &  292,308 & 284,695      \\ \bottomrule
\end{tabular}}
\label{tab dataset}
\end{table}

\section{Baselines}
\label{app: baseline}
\begin{enumerate}[leftmargin=0.5cm]
    \item Collaborative filtering based (CF-SRec)
    \begin{itemize} [leftmargin=*,itemsep=0pt, topsep=0pt]
        \item \textbf{GRU4Rec} \cite{hidasi2015session} employs a recurrent neural network (RNN) to capture user behavior sequences for session-based recommendation.
        \item \textbf{BERT4Rec} \cite{sun2019bert4rec} utilizes bidirectional self-attention mechanisms and a masked item prediction objective to model complex user preferences from interaction sequences.
        \item \textbf{NextItNet} \cite{yuan2019simple} applies temporal convolutional layers to capture both short-term and long-term user preferences.
        \item \textbf{SASRec} \cite{kang2018self} is a self-attention based recommender system designed to capture long-term user preference.
    \end{itemize}

    \item Language model based (LM-based)
    \begin{itemize} [leftmargin=*,itemsep=0pt, topsep=0pt]
        \item \textbf{CTRL} \cite{li2023ctrl} initializes the item embeddings of the backbone recommendation models with textual semantic embeddings using the RoBERTa \cite{liu2019roberta} encoding models. And fine-tunes the backbone models for the recommendation task. 
        \item \textbf{RECFORMER} \cite{li2023text} leverages a Transformer-based framework for sequential recommendation, representing items as sentences by flattening the item title and attributes.
    \end{itemize}
    \item Large Language Model based (LLM4Rec)
        \begin{itemize} [leftmargin=*,itemsep=0pt, topsep=0pt]
        \item \textbf{TALLRec} \cite{bao2023tallrec} fine-tunes LLMs for the recommendation task by formulating the recommendation task as a target item title generation task.
        \item \textbf{LLaRA} \cite{10.1145/3626772.3657690} uses CF-SRec to incorporate behavioral patterns into LLM. To align the behavioral representations from the CF-SRec this model employs a hybrid prompting which is a concatenated form of textual embedding and item representations.
        \item \textbf{CoLLM} \cite{10.1145/3637528.3671931} integrates the collaborative information as a distinct modality into LLMs by extracting and injecting item embeddings from CF-SRec. 
        \item \textbf{A-LLMRec} \cite{10.1145/3637528.3671931} enables LLMs to leverage the CF knowledge from CF-SRec and item semantic information through a two-stage learning framework. 
    \end{itemize}
\end{enumerate}

\section{Implementation Details}
\label{app: implementation details}
In our experiments, we adopt SASRec as a CF-SRec backbone for CoLLM, LLaRA, A-LLMRec, and ~\proposed, with its item embedding dimension fixed to 64 and batch size set to 128.
For LLM4Rec baselines, including Stage-2 of A-LLMRec, the batch size is 20 for the Movies, Scientific, and CDs datasets, while 16 is used for the Electronics dataset. For Stage-1 of A-LLMRec, the batch size is set to 64.
When using Intel Gaudi v2, we set the batch size to 4 due to 8-bit quantization constraints.
All LLM4Rec models are trained for a maximum of 10 epochs, with validation scores evaluated at every 10\% of the training progress within each epoch, where early stop with patience of 10 is applied to prevent over-fitting.
All models are optimized using Adam with a learning rate 0.0001, and the dimension size of the projected embedding $d'$ is 128.
Experiments are conducted using a single NVIDIA GeForce A6000 (48GB) GPU and a single Gaudi v2 (100GB).

\section{Additional Experiments}

\subsection{Effectiveness of Input Prompts}
\label{app: prompt study}
Note that rather than explicitly having the user representations in the input prompt, we rely on the [UserOut] token to extract the user representations as shown in (Table \ref{tab next item retrieval prompt} (b)).
In Table \ref{tab prompt study}, to validate whether it is sufficient, we compare the performance of~\proposed~with and without the explicit user representations.
The results show a comparable performance between the two prompts. This suggests that through Equation \ref{Eq distill}, the sequential information contained in the user representation is effectively transferred to the LLMs, enabling them to understand sequential dependencies using only the user interaction sequence without explicitly incorporating the user representation in the prompt. Furthermore, omitting the user representation and its associated text from the prompt reduces input prompt length, improving training/inference efficiency, which implies the practicality of \proposed's prompt.

\begin{table}[t]
\caption{Performance comparison of prompts with/without explicit user representations.}
\vspace{-2ex}
\resizebox{0.7\linewidth}{!}{
\begin{tabular}{c|c||c||c}
\toprule
Dataset                      & Metric  & \makecell{With User \\Representations} & \proposed \\ \midrule\midrule
\multirow{2}{*}{Movies}      & NDCG@10 & \textbf{0.3625}    & 0.3560                   \\ \cline{2-4} 
                             & HR@10   & \textbf{0.5626}    & 0.5569                   \\ \midrule\midrule
\multirow{2}{*}{Scientific}  & NDCG@10 & 0.3342    & \textbf{0.3388}                   \\ \cline{2-4} 
                             & HR@10   & 0.5516    & \textbf{0.5532}                   \\\midrule\midrule
\multirow{2}{*}{Electronics} & NDCG@10 & 0.2924    & \textbf{0.3044}                   \\ \cline{2-4} 
                             & HR@10   & 0.4725    & \textbf{0.4885}                   \\ \bottomrule
\end{tabular}}
\vspace{-2.5ex}
\label{tab prompt study}
\end{table}

\subsection{Distillation with Contrastive Learning}
\label{app: contra distillation}
Recall that we distill sequential information from CF-SRec to LLMs using MSE loss in Equation \ref{Eq distill}. To further investigate the impact of the distillation loss function, we adapt a naive contrastive learning method for sequential information distillation, i.e., Equation \ref{Eq distill-contra}.

\begin{equation}
    \mathcal{L}_{\text{Distill-Contrastive}} = -\underset{u \in \mathcal{U}}{\mathbb{E}}\text{log}\frac{e^{s(f_\mathit{user}(\mathbf{h}_{\mathcal{U}}^u),f_\mathit{CF-user}(\mathbf{O}_u))}}{\sum_{k\in\mathcal{U}} e^{s(f_\mathit{user}(\mathbf{h}_{\mathcal{U}}^u),f_\mathit{CF-user}(\mathbf{O}_k))}}
    \label{Eq distill-contra}
\end{equation}
Table \ref{tab contrastive} shows the performance of different distillation loss functions, and we have the following observations: 1) MSE loss (Equation \ref{Eq distill}) consistently outperforms contrastive loss (Equation \ref{Eq distill-contra}) across all datasets, indicating that effective sequential information transfer to LLMs requires more than just aligning overall trends. Instead, explicitly matching fine-grained details in representations plays a crucial role in preserving sequential dependencies. 2) Performance degradation occurs when inference is performed on shuffled sequences regardless of the chosen loss function, indicating that both losses successfully captures the sequential information.

\begin{table}[]
\caption{Distillation with contrastive learning (NDCG@10).}
\resizebox{0.8\linewidth}{!}{
\begin{tabular}{c|c||c||c||c}
\toprule
Distillation Loss            & Inference  & Movies                                                                         & Scientific                                                                      & Electronics                                                                     \\ \midrule\midrule
\multirow{3}{*}{Contrastive} & Original          & 0.3410  & 0.2767  & 0.2553 \\  
 & Shuffle           & 0.3151 & 0.2638         & 0.2398     \\ \cmidrule{2-5} 
 & Change ratio      & (-7.60\%)    & (-4.66\%)    & (-6.07\%)  \\ \midrule\midrule
\multirow{3}{*}{\proposed~(MSE)}    & Original          & \textbf{0.3560}  & \textbf{0.3388}  & \textbf{0.3044} \\ 
 & Shuffle           & 0.3272 & 0.3232    & 0.2845  \\ \cmidrule{2-5} 
 & Change ratio      & (-8.10\%)  & (-4.60\%)  & (-6.53\%) \\ \bottomrule
\end{tabular}}
\label{tab contrastive}
\vspace{-1.5ex}
\end{table}



\begin{figure}[t]
    \centering
    \includegraphics[width=0.925\linewidth]{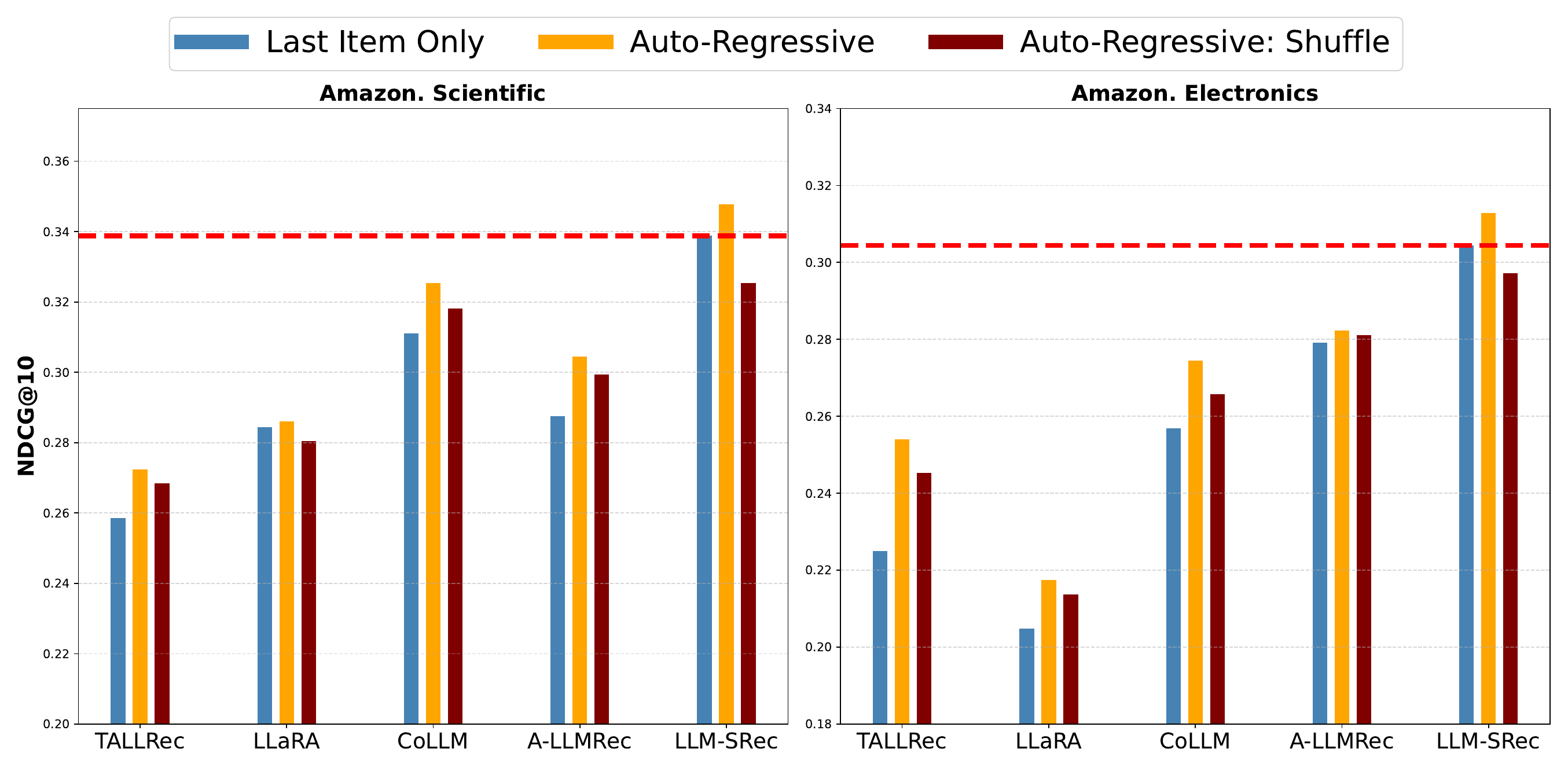}
    \vspace{-2ex}
    \caption
    {Performance with auto-regressive training strategy.}
    \label{fig: auto-regressive}
    \vspace{-1ex}
\end{figure}
\subsection{Auto-regressive Training}
\label{app: autoregressive}
Recall that, for training efficiency, we only consider the last item in the user sequences as the target item to train~\proposed. On the other hand, we can consider all items in the user sequence as the target item to train the models in an auto-regressive manner.
As shown in Figure \ref{fig: auto-regressive}, when all the models are trained in an auto-regressive manner, their performance improves, demonstrating the benefits of leveraging more historical interactions.
One notable result is that our ~\proposed~without the auto-regressive training outperforms other models with the auto-regressive strategy. This is a notable result as the number of samples used for training is much less without auto-regressive training. This result underscores the efficacy of our framework in capturing sequential patterns.
{Furthermore, in the shuffled setting, baselines exhibit a relatively small change ratio compared to ~\proposed, indicating that they still fall short of understanding sequence although the baselines learn the fine-grained item sequences through the auto-regressive manner.}

\end{document}